\documentclass{article}

\PassOptionsToPackage{numbers, compress}{natbib}

\usepackage[preprint]{neurips_2025}

\usepackage[utf8]{inputenc} 
\usepackage[T1]{fontenc}    
\usepackage{hyperref}       
\usepackage{url}            
\usepackage{booktabs}       
\usepackage{amsfonts}       
\usepackage{nicefrac}       
\usepackage{microtype}      
\usepackage{xcolor}         
\usepackage{graphicx}       
\usepackage{amsmath}        
\usepackage{multirow}
\usepackage{utfsym}
\usepackage{makecell}
\usepackage{colortbl}
\usepackage{CJKutf8}

\title{ClinBench-HPB: A Clinical Benchmark for Evaluating LLMs in Hepato-Pancreato-Biliary Diseases}

\author{
    Yuchong Li$^{1,2,^{\star}}$,   Xiaojun Zeng$^{2,3^{\star}}$,  Chihua Fang$^{3}$, Jian Yang$^{3}$, Fucang Jia$^{2,4,^{\dagger}}$, Lei Zhang$^{1,^{\dagger}}$\thanks{\small $^{\star}$Equal contribution. $^{\dagger}$Corresponding author} \\
    $^{1}$The Hong Kong Polytechnic University \\ $^{2}$Shenzhen Institutes of Advanced Technology, CAS\\ $^{3}$Zhujiang Hospital, Southern Medical University \\ $^{4}$The Key Laboratory of Biomedical Imaging Science and System, CAS \\
    \small yuchong.li@connect.polyu.hk, smuzxj@hotmail.com, fc.jia@siat.ac.cn, cslzhang@comp.polyu.edu.hk \\
}

\makeatletter
\def\thanks#1{\protected@xdef\@thanks{\@thanks
        \protect\footnotetext{#1}}}
\makeatother

\begin{document}

\maketitle

\begin{abstract}
Hepato-pancreato-biliary (HPB) disorders represent a global public health challenge due to their high morbidity and mortality. Although large language models (LLMs) have shown promising performance in general medical question-answering tasks, the current evaluation benchmarks are mostly derived from standardized examinations or manually designed questions, lacking HPB coverage and clinical cases.
To address these issues, we systematically eatablish an HPB disease evaluation benchmark comprising 3,535 closed-ended multiple-choice questions and 337 open-ended real diagnosis cases, which encompasses all the 33 main categories and 465 subcategories of HPB diseases defined in the International Statistical Classification of Diseases, $10^{th}$ Revision (ICD-10). 
The multiple-choice questions are curated from public datasets and synthesized data, and the clinical cases are collected from prestigious medical journals, case-sharing platforms, and collaborating hospitals.
By evalauting commercial and open-source general and medical LLMs on our established benchmark, namely ClinBench-HBP, we find that while commercial LLMs perform competently on medical exam questions, they exhibit substantial performance degradation on HPB diagnosis tasks, especially on complex, inpatient clinical cases. Those medical LLMs also show limited generalizability to HPB diseases. 
Our results reveal the critical limitations of current LLMs in the domain of HPB diseases, underscoring the imperative need for future medical LLMs to handle real, complex clinical diagnostics rather than simple medical exam questions. The benchmark will be released at \href{https://clinbench-hpb.github.io/}{https://clinbench-hpb.github.io}.
\end{abstract}

\section{Introduction}

\begin{table}[htbp]
  \centering
  \caption{Comparison between ClinBench-HPB and existing studies for evaluating LLMs in HPB diseases. \#Dis., \#Ques., and \#LLMs denote the numbers of HPB disease categories, evaluation questions, and assessed LLMs, respectively. 
"Know." and "Real Patient" indicate whether or not the evaluation set includes disease knowledge or real-patient diagnostic data, respectively. (HBV: hepatitis B virus; HCV: hepatitis C virus; HCC: hepatocellular carcinoma).}
  \scalebox{0.9}{
    \begin{tabular}{clccccc}
    \toprule
    \multicolumn{2}{c}{Task} & \#Dis. & \#Ques. & \#LLMs & Know. & Real Patient \\
    \midrule
    \multicolumn{1}{c}{\multirow{4}[2]{*}{\makecell[c]{Expert-\\driven}}} &   Cirrhosis and HCC \cite{yeo2023assessing}  & 2     & 164    & 1  &  \checkmark & \scalebox{0.75}{\usym{2613}}\\
          &   Acute pancreatitis \cite{du2024exploring}   & 1     & 91    & 2  &\checkmark & \scalebox{0.75}{\usym{2613}}\\
          &   Incidental hepatic steatosis \cite{vong2025automated}  & 1 & 200 & 3 & \checkmark  & \scalebox{0.75}{\usym{2613}} \\
          &  Liver transplantation \cite{endo2023quality} & 1 & 493 & 1 &  \checkmark & \scalebox{0.75}{\usym{2613}} \\
    \midrule
    \multicolumn{1}{c}{\multirow{3}[2]{*}{\makecell[c]{RAG-\\based}}} &  HCV management \cite{kresevic2024optimization}     & 1 & 20 & 1 & \checkmark & \scalebox{0.75}{\usym{2613}} \\
          &    HBV therapy \& HCC monitoring \cite{ge2024development}   & 2     & 10    & 1 &  \checkmark & \scalebox{0.75}{\usym{2613}}\\
          &   Incidental hepatobiliary findings \cite{dietrich2025evaluating}& 1     & 319   & 2 & \checkmark & \scalebox{0.75}{\usym{2613}}\\
    \midrule
    Ours  & ClinBench-HPB & 33+465 & 3535+337 & 26  & \checkmark & \checkmark \\
    \bottomrule
    \end{tabular}}%
  \label{tab:1}%
  \vspace{-4pt}
\end{table}%

Hepato-pancreato-biliary (HPB) diseases, which affect the liver, pancreas, and biliary tract, manifest a broad clinical spectrum due to unique dual endocrine/exocrine functions of these organs, ranging from inflammatory disorders (\textit{e.g.}, hepatitis) to malignancies (\textit{e.g.}, pancreatic adenocarcinoma).
These diseases pose significant global health challenges due to their high incidence and mortality rates \cite{hpb2,hpb1}.
The diagnosis of HPB diseases requires a comprehensive analysis of multi-scource unstructured data from patient complaints, medical history, laboratory tests, and imaging studies. It also relies on individual knowledge and experience of physicians, which can lead to anchoring or availability bias for diagnostic errors \cite{dahiya2025large}. 
Furthermore, extracting information from extensive electronic health records and clinical documentation is tedious and time-consuming, which significantly impacts diagnostic efficiency and increases physician workload \cite{gong2024large, li2024lkan}.
With the remarkable progress of large language models (LLMs) \cite{qwen2.5, llama3.1, achiam2023gpt-4, ouyang2022training} in recent years, it has been becoming popular to employ LLMs to summarize medical texts and identify abnormalities, thus enhancing diagnostic efficiency and supporting clinical decision-making \cite{dahiya2025large, gong2024large, li2024lkan, giuffre2024optimizing}.

To facilitate the application of LLMs for medical diagnosis, one critical issue is how to evaluate the diganostic performance of LLMs. As shown in Table \ref{tab:1}, current research on the evaluation of LLMs in HPB diseases can be categorized into two categories.
One category employs an expert-driven evaluation paradigm, where clinicians design disease-specific questions to assess knowledge, diagnostic reasoning, and patient communication \cite{yeo2023assessing, du2024exploring, vong2025automated, endo2023quality, geevarghese2025extraction, gorelik2024using, pugliese2024accuracy, cao2023accuracy, goglia2025using, mao2025a}.
Another category focuses on LLMs' knowledge retrieval and integration capability. Leveraging authoritative guidelines and retrieval-augmented generation (RAG), researchers assess the accuracy and evidence reliability of LLMs by answering clinically relevant queries \cite{kresevic2024optimization,ge2024development,dietrich2025evaluating,wu2024diagnosis}.

However, the existing studies on evaluating LLM in HPB diseases face two critical limitations.
First, their coverage of HPB diseases is incomplete and lacks clinical relevance. According to the International Classification of Diseases, $10^{th}$ Revision (ICD-10) \cite{icd-10-zh, icd-10-en}, there are 33 main categories and 465 subcategories for HPB diseases. However, existing HPB evaluation benchmarks only cover several of these diseases, and the assessments are based on a few questions hand-crafted with short contexts.
This gap hinders an accurate assessment of LLMs' applications in HPB disgnosis.
Secondly, existing open-ended evaluation metrics cannot assess HPB clinical diagnostic problems.
Current research mainly employs three types of automated assessment methods: traditional NLP metrics (\textit{e.g.}, ROUGE, BLEU) \cite{PediatricsGPT,26,91,102}, BERT-based semantic similarity \cite{103,105,medic}, and multi-dimensional LLM scoring \cite{PediatricsGPT,PediaBench, llm_eval1,llm_eval2}. 
Unfortunately, traditional NLP metrics fail to capture medical term synonymy (\textit{e.g.}, matching between "postoperative status" and "prognostic condition"). BERT-based embeddings fail to distinguish nuanced name differences (\textit{e.g.}, "acute cholecystitis" and "acute cholangitis" are distinct conditions, yet produce a high BERTscore). LLM automated scoring is vulnerable to complex cases with multiple co-existing conditions.

\begin{figure}
    \centering
    \includegraphics[width=1\linewidth]{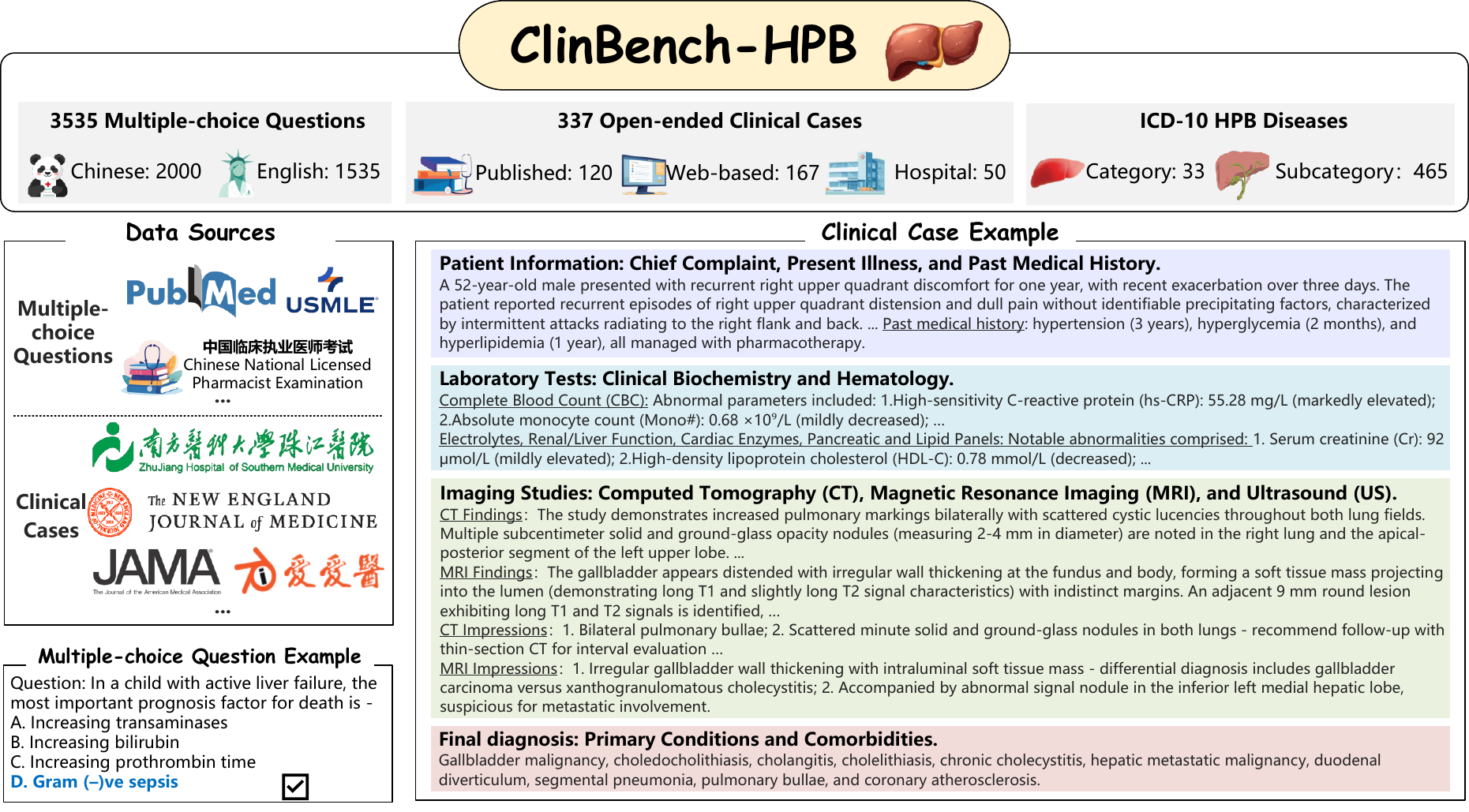}
    \caption{Overview of our established ClinBench-HPB. The benchmark comprises 3,535 multiple-choice questions and 337 real-world clinical cases, covering 33 main categories and 465 subcategories of HPB diseases defined in ICD-10.}
    \label{fig:1}
    \vspace{-2pt}
\end{figure}

To address the above-mentioned issues, we propose ClinBench-HPB, a clinically oriented benchmark across the spectrum of HPB diseases defined under ICD-10.
As shown in Fig. \ref{fig:1}, the benchmark comprises 3,535 closed-ended multiple-choice questions and 337 open-ended diagnostic cases.
The multiple-choice questions are designed to assess LLMs' knowledge coverage. We filter HPB-related questions from public general medical evaluation benchmarks and synthesize additional questions for rare diseases to ensure completeness.
Diagnostic cases are sourced from authoritative medical journals, case-sharing platforms, and collaborative hospitals to evaluate LLMs' practical diagnostic capabilities.
For each case, the model is required to analyze patient information, laboratory results, and imaging studies to generate detailed diagnoses.
We introduce disease-level and patient-level recall metrics using LLM-as-a-judge. 
The evaluation process involves a strict assessment and a check step that assesses the inclusion of reference diseases in model predictions along six dimensions: anatomical specificity, pathological mechanism, etiological origin, temporal characteristics, test identification, and treatment strategy.

We inclusively evaluate the performance of commercial, open-source general-purpose, medical and reasoning-enhanced LLMs using the benchmark.
Extensive experiments show that although current commercial LLMs achieve high accuracy on multiple-choice questions derived from medical licensing examinations, they remain inadequate for real-world clinical case diagnosis. 
Open-source general-purpose LLMs are approaching commercial models, while their analytical reasoning capabilities on complex tasks still require further improvment.
Medical models exhibit performance degradation on out-of-distribution HPB data, even underperforming baseline models on certain clinical diagnosis cases. 
The reasoning-enhanced LLMs can more effectively leverage the medical knowledge, demonstrating the potential to address complex diagnostic tasks.
Our evaluation benchmark reveals critical limitations of current LLMs in the HPB disease domain, underscoring the need to focus on complex real-world clinical diagnostics rather than simplistic medical examination questions.

\section{Related Work}

\textbf{General-Purpose LLM for HPB Diseases}.
Previous studies mostly evaluate the LLMs in specific HPB diseases and clinical scenarios, including liver \cite{yeo2023assessing, vong2025automated, endo2023quality, geevarghese2025extraction, pugliese2024accuracy, cao2023accuracy}, biliary \cite{gorelik2024using, mao2025a}, and pancreatic diseases \cite{du2024exploring, goglia2025using, bhayana}.
Yeo et al. \cite{yeo2023assessing} assessed the response quality of ChatGPT \cite{ouyang2022training} using a set of 164 clinical questions collected from institutional questionnaires and online communities. Each question was processed twice, and the responses were independently evaluated by three experts.
Du et al. \cite{du2024exploring} constructed a test set comprising 18 subjective and 73 objective questions derived from acute pancreatitis guidelines and public databases. They evaluated ChatGPT and GPT-4 \cite{achiam2023gpt-4} by three physicians.
These studies demonstrated that while LLMs possess basic medical knowledge, their responses tend to be overly generic, lack precision in diagnostic thresholds, and exhibit information latency.
Subsequent studies have integrated LLMs with RAG techniques \cite{kresevic2024optimization, ge2024development, dietrich2025evaluating, wu2024diagnosis}.
Ge et al. \cite{ge2024development} built a vector database using 30 guidelines from the American Association for the Study of Liver Diseases (AASLD), improving response relevance through similarity-based context retrieval using GPT-3.5-turbo and GPT-4.
Kresevic et al. \cite{kresevic2024optimization} optimized GPT-4 Turbo via context augmentation, data cleaning, tabular reformatting, and prompt engineering, achieving incremental accuracy improvements on 20 Hepatitis C related questions. 
However, as shown in Table \ref{tab:1}, existing studies evaluate LLMs using a very limited number of HPB disease categories, questions, and models, and assess model knowledge using manually crafted questions. 

\begin{figure}
    \centering
    \includegraphics[width=1\linewidth]{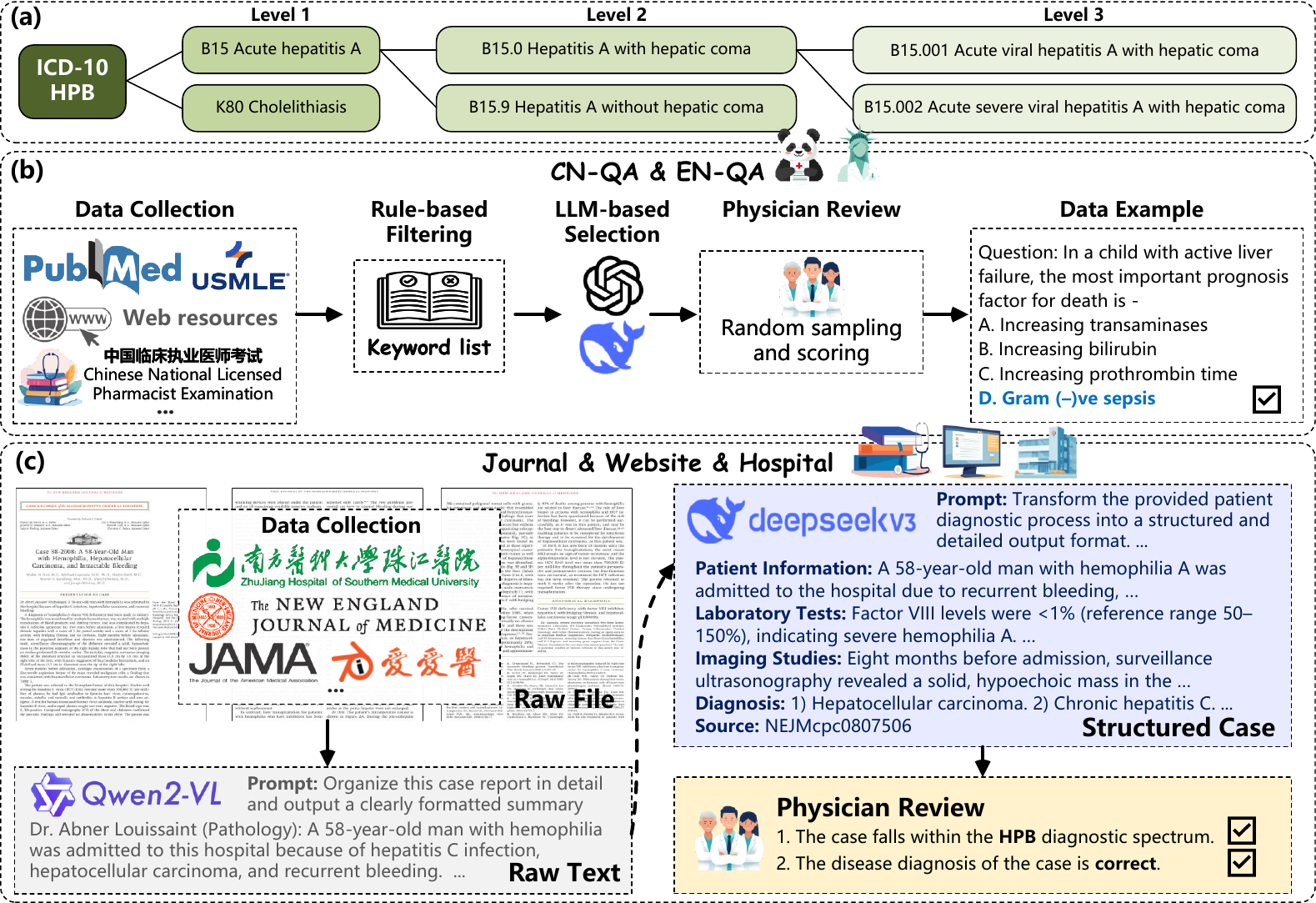}
    \caption{Overview of the ClinBench-HPB construction process. (a) Three-level organization of HPB disorders in ICD-10 with representative examples. (b) The data collection and processing pipeline of multiple-choice questions. (c) Clinical cases guided by the disease classification.}
    \label{fig:2}
\end{figure}

\textbf{Medical LLMs and Evaluation Benchmarks}.
The development of medical LLMs improves the accuracy of professional question-answering and provides clinical decision support in healthcare. 
The training strategy includes domain enhancement of general LLMs (\textit{e.g.}, Qwen \cite{qwen2.5} and LLaMA series \cite{llama3.1}) using supervised fine-tuning (SFT) and reinforcement learning (RL) \cite{ultramedical, openbiollm, o1-part3, m1, huatuo-o1}.
HuatuoGPT-o1 \cite{huatuo-o1} fine-tunes multiple general LLMs via SFT and proximal policy optimization (PPO), leveraging 40K synthetic  samples to improve medical reasoning.
Baichuan-M1 \cite{m1} trains a 14B medical LLM from scratch on a high-medical-proportion corpus, adapting 72B general-purpose LLMs in downstream tasks.
To assess the medical capabilities of LLMs, existing benchmarks mainly consist of medical licensing exam questions \cite{medqa, medmcqa, cmb, medbullets}, question-answering pairs extracted from PubMed publications \cite{pubmedqa}, and medical subsets of general-purpose benchmarks \cite{mmlu-pro, gpqa}.
MedMCQA\cite{medmcqa} contains over 6K test questions from All India Institute of Medical Sciences (AIIMS PG) and National Eligibility cum Entrance Test (NEET PG), covering 21 medical subjects and 2,400 healthcare topics. 
PubMedQA \cite{pubmedqa} transforms PubMed abstracts into 1,000 expert-annotated multiple-choice questions with yes/no/maybe options, using other sections as contextual support.
However, current benchmarks primarily assess knowledge through standardized medical examination questions, which exhibit significant gaps to real-world diagnostic scenarios. In addition, the closed-ended multiple-choice questions rely heavily on heuristic reasoning patterns \cite{nejm_ai}, which cannot adequately assess the applicability of LLMs to practical clinical cases.

\section{Construction of ClinBench-HBP}

The construction process of ClinBench-HBP is illustrated in Fig. \ref{fig:2}. 
We extract all categories of HPB diseases from ICD-10, then collect 3,535 closed-ended multiple-choice questions and 337 open-ended clinical cases with their composition detailed in Table \ref{tab:2}, enabling a comprehensive evaluation of LLM in HPB diseases. In particular, through the incorporation of complex public and proprietary cases, we significantly enhance the clinical relevance of ClinBench-HBP.

\begin{table}[htbp]
  \centering
  \caption{Statistics of the five subsets in ClinBench-HPB. \#Size, \#Avg Lens, \#Avg Ans denote the number of questions, average token length per question (using Qwen \cite{qwen2.5} tokenizer), and average number of answers per question, respectively.}
    \begin{tabular}{lcccccc}
    \toprule
    Subset & \# Size & \# Avg Lens & \# Avg Ans & Data Source & Measure & Language \\
    \midrule
    CN-QA  & 2000  &  87.08     & 1.20   &   Chinese medical exams    &   Accuracy    & CN \\
    EN-QA  & 1535  &  73.99     & 1.00     &   English medical exams    &   Accuracy    & EN \\
    Journal & 120   &  357.79     &  1.46    &   Medical journals    &   Recall    & EN\&CN \\
    Website & 167   &   479.99    &   3.68    &    Case-sharing websites   &    Recall   & CN \\
    Hospital & 50    &  1389.74     &  7.86     &   Collaborative hospital    &   Recall    & CN \\
    \bottomrule
    \end{tabular}%
  \label{tab:2}%
  \vspace{-5pt}
\end{table}%

\subsection{Data Collection and Processing}
\vspace{-3pt}

We manually screen disease terms related to HPB diseases from ICD-10, excluding entries with ambiguous references (\textit{e.g.}, "A18.8 Tuberculosis of other specified organs").
Following the coding system \cite{icd-10-zh}, we categorize these diseases into three tiers, as shown in Fig. \ref{fig:2}(a). Our evaluation benchmark consists of 33 main categories (Level 1) and 465 valid subcategories (Levels 2 and 3).
For the multiple-choice section, we select relevant questions from established medical datasets \cite{medqa, medmcqa, cmb, pubmedqa, mmlu-pro} and supplement them with synthetic data targeting ICD-10-coded rare diseases.
These questions are organized into two categories: CN-QA for Chinese-language items and EN-QA for English-language items.
The clinical diagnosis section includes cases sourced from medical journals (Journal), case-sharing websites (Website), and collaborating hospitals (Hospital). All cases undergo standardized structuring and are validated by physicians to ensure diagnostic accuracy.

\textbf{Chinese (CN) / English (EN) Medical Examination Questions.} 
The construction process of this subset is illustrated in Fig. \ref{fig:2}(b).
The multiple-choice questions are designed to assess LLMs' domain-specific knowledge. 
We first aggregate existing medical examination-based test sets and the medical subsets from general evaluation benchmarks, including CMB (11,200), MedQA-English (1,273), MedMCQA (4,183), PubMedQA (500), and MMLU-Pro (1,535).
These questions cover content from the United States Medical Licensing Examination (USMLE), the Chinese National Licensed Pharmacist Examination, and PubMed abstracts, among others. We further supplement approximately 1,000 Chinese medical practice questions using "hepatobiliary surgery" as a keyword from online resources. 
In addition, we categorize ICD-10-HPB diseases by level-2 groupings and employ an internet-connected LLM to gather relevant materials for each group. 
The LLM is then used to convert these materials into multiple-choice questions, expanding the coverage of rare diseases. 
To reduce hallucinations, we constrain question generation to strictly follow the retrieved source material, yielding approximately 1,400 synthesized questions.

The data cleaning process consists of two stages: rule-based filtering and LLM-assisted selection.
For Chinese items, we construct keyword lists for retention and removal through random data analysis.
We retain questions containing at least one domain-relevant keyword ("liver", "biliary", or "pancreas")  while filtering out those containing clinically irrelevant terms (\textit{e.g.}, "recommendations", "association").
After this, we perform an LLM-based selection.
We employ GPT-4o \cite{achiam2023gpt-4} and DeepSeek-V3 \cite{deepseekv3} to perform dual-validation classification of each question's disease domain, categorizing them into hepatic diseases, biliary disorders, pancreatic pathologies, portal hypertension, or others (with non-exclusive categorization). 
Only questions classified as HPB-related by both models are retained for the final dataset.
For the synthetic section, each question is evaluated by DeepSeek-V3 using a 5-point clinical relevance scale with justification. 
Questions scoring 1-2 are excluded, and those scoring 3 undergo model-guided revision. Questions rating 4-5 are retained without modification.
From the 1,362 questions generated, we randomly select 500 items for quality assessment by four licensed physicians. 
More than 96\% of the questions meet quality standards, with an agreement of 94\% among physicians.
Finally, the subset of multiple-choice questions includes 2,000 Chinese (CN-QA) and 1,535 English (EN-QA) items.
The CN-QA section comprises 1,820 single-answer and 180 multiple-answer questions, and the EN-QA portion contains 173 items from public datasets and 1,362 clinically validated synthetic entries, all with single answers.

\textbf{Clinical Cases.}
The construction process of this subset is illustrated in Fig. \ref{fig:2}(c).
We first collect HPB cases from publicly available case reports and case-sharing websites.
The case reports are primarily sourced from the JAMA Clinical Challenge (JAMA) and NEJM Case Report (NEJM) published between 2000 and 2025.
The JAMA dataset presents real-world clinical cases through brief vignettes, each followed by a multiple-choice question to identify the most likely diagnosis, determining the next diagnostic step, or selecting the optimal management strategy.
To standardize the format and enhance difficulty, we change them into open-ended forms by retaining the original questions and eliminating the options, resulting in 69 cases.
The NEJM dataset features 32 case records of the Massachusetts General Hospital, including detailed patient presentations, comprehensive discussions of diagnostic reasoning, differential diagnoses, and management decisions.
To enhance disease diversity, we supplement 19 rare HPB cases from medical journals indexed in Google Scholar and China National Knowledge Infrastructure (CNKI).
In addition, we screen HPB cases from open-access case-sharing websites. 
To ensure data quality, we exclusively select cases that contain at least one laboratory examination or imaging study result, along with a preliminary or definitive diagnosis, obtaining 167 cases. 
To further evaluate the model's ability to diagnose HPB diseases in inpatient clinical cases, we collect 50 detailed inpatient cases from the collaborative hospital. 
As illustrated in Fig. \ref{fig:1} and Table \ref{tab:2}, the hospital-collected cases exhibit longer contextual information and a higher average number of comorbidities compared to other sources. 
These real-world patient data effectively reflect the complexity of actual clinical practice, thereby enabling an objective assessment of LLMs' genuine effectiveness in clinical tasks.
During the data preprocessing phase, we employ a locally deployed Qwen2-VL-72B \cite{qwen2vl} to convert the original medical case reports into textual format. 
After manual check to remove protected health information, the textual data are structured using DeepSeek-V3 \cite{deepseekv3}.
As shown in Fig. \ref{fig:1}, each medical case consists of four key components:

\textbf{Patient Information}, including chief complaints, medical history, and family history. This section provides essential background for understanding the patient's condition.

\textbf{Laboratory Tests}, such as blood tests and biochemical analyses, provide objective data to support clinical assessment.

\textbf{Imaging Studies}, including CT, MRI, and ultrasound. In hospital records, each imaging report contains both "Findings" (detailed observations by radiologists) and "Impressions" (diagnostic interpretations), which aid in disease localization and characterization.

\textbf{Final Diagnosis}, which encompasses both primary conditions (\textit{e.g.}, hepatocellular carcinoma) and comorbidities (\textit{e.g.}, hypertension), reflecting the comprehensive clinical evaluation.

We invite four licensed physicians to validate the case diagnosis, with each receiving a \$50 honorarium. 
The final cohort includes 337 clinically validated cases (Journal: 120, Website: 167, Hospital: 50) encompassing 33 major HPB categories within the ICD-10 classification system.

\subsection{Evaluation Method}
\vspace{-3pt}

\textbf{Objective Questions}.
For closed-ended multiple-choice questions, we evaluate the performance of LLMs using accuracy metrics, with the prompt explicitly specifying the output format (\textit{e.g.}, "Answer: A" for single-choice and "Answer: ABC" for multiple-choice). The prompts we used can be found in \textbf{Appendix \ref{appendix1.1}}. Responses are processed by normalizing case and punctuation and then extracting answers through regular expression matching of alphanumeric characters following "Answer". 
For non-compliant responses, GPT-4o-mini \cite{gpt4omini} is used to analyze and extract option letters. 
To mitigate LLM's sensitivity to option positions \cite{zheng2023large}, we perform $min(4,n-1)$ circular right shifts on its options, where $n$ is the number of options.
Given the original option sequence $O=[o_1, o_2, ..., o_n]$, the shifted sequence $O^{(k)}$ after $k$ shifts is defined as:
\[o^{(k)}_i=o_{(i-k-1)\mod n+1},  i\in\{1,2,...,n\}, k\in\{1,2,...,min(4,n-1)\}. \]

The model predicts after each shift, producing $N\in\{2,3,4\}$ outputs per question. The mean and standard deviation (std) of these predictions are then computed to assess model performance.


\begin{table}[htbp]
  \centering
  \caption{Results of different LLMs on ClinBench-HPB. The best-performing model in each group is \textbf{in-bold}, and the second best is \underline{underlined}. Categories: Comm (commercial), Open (open-source general-purpose), Med (medical), Reason (reasoning-enhanced). Metrics: Pt (patient-level recall), Dis (disease-level recall), $Avg_{q,p,d}$ (quantity-weighted average), $Avg$ (arithmetic mean of $Avg_q$ and $Avg_p$). "$\dagger$" means the result on sampled subset.}
\resizebox{1.0\textwidth}{!}{
    \begin{tabular}{cl|cccccccccccc}
    \toprule
    \multicolumn{1}{c}{\multirow{3}[2]{*}{Category}} & \multicolumn{1}{c|}{\multirow{3}[2]{*}{Model}} & \multicolumn{3}{c}{Multiple Choice} & \multicolumn{8}{c}{Case Diagnosis}                            & \multirow{3}[2]{*}{$Avg$} \\
    \cmidrule(r){3-5} \cmidrule(r){6-13}
          &       & \multicolumn{1}{c}{\multirow{2}[1]{*}{\makecell[c]{CN-QA\\(2,000)}}} & \multicolumn{1}{c}{\multirow{2}[1]{*}{\makecell[c]{EN-QA\\(1,535)}}} & \multirow{2}[1]{*}{$Avg_q$} & \multicolumn{2}{c}{Journal} & \multicolumn{2}{c}{Website} & \multicolumn{2}{c}{Hostipal} & \multirow{2}[1]{*}{$Avg_p$} & \multirow{2}[1]{*}{$Avg_d$} &  \\
          &       &       &       &       & Pt(120) & Dis(175) & Pt(167) & Dis(614) & Pt(50) & Dis(393) &       &       &  \\
    \midrule
        \multicolumn{1}{c}{\multirow{9}[2]{*}{Comm}} & GPT-4o & 0.594  & 0.889  & 0.722  & 0.319  & 0.443  & 0.379  & 0.720  & 0.090  & 0.598  & 0.315  & 0.639  & 0.518  \\
          & OpenAI-o1 & $0.760^\dagger$  & $0.917^\dagger$  & $0.828^\dagger$  & 0.490  & 0.587  & 0.525  & 0.802  & 0.110  & 0.641  & 0.451  & 0.717  & $0.639^\dagger$  \\
          & OpenAI-o3mini & 0.666  & 0.894  & 0.765  & \textbf{0.558} & \textbf{0.637} & 0.539  & 0.805  & 0.115  & 0.641  & \textbf{0.483} & 0.725  & 0.624  \\
          & Claude3.5-sonnet & 0.623  & 0.878  & 0.734  & 0.394  & 0.514  & \underline{0.546}  & \textbf{0.828} & 0.050  & 0.570  & 0.418  & 0.696  & 0.576  \\
          & Gemini2.5-pro & 0.724  & \textbf{0.897} & 0.799  & \underline{0.500}  & \underline{0.603}  & 0.504  & 0.809  & 0.135  & 0.712  & \underline{0.448}  & 0.747  & 0.624  \\
          & Qwen2.5-Max & 0.687  & 0.876  & 0.769  & 0.329  & 0.460  & 0.421  & 0.762  & 0.130  & 0.687  & 0.345  & 0.692  & 0.557  \\
          & DeepSeekV3-1226 & 0.682  & 0.881  & 0.769  & 0.327  & 0.476  & \textbf{0.548} & \underline{0.826}  & \textbf{0.220} & \textbf{0.740} & 0.421  & \textbf{0.745} & 0.595  \\
          & DeepSeekV3-0324 & \textbf{0.803} & 0.887  & \underline{0.839} & 0.381  & 0.513  & 0.527  & 0.809  & \underline{0.175}  & \underline{0.716}  & 0.423  & \underline{0.735}  & \underline{0.631}  \\
          & DeepSeek-R1 & \underline{0.799}  & \underline{0.893}  & \textbf{0.840}  & 0.433  & 0.553  & 0.512  & 0.785  & 0.108  & 0.648  & 0.424  & 0.705  & \textbf{0.632} \\
    \midrule
        \multicolumn{1}{c}{\multirow{6}[2]{*}{Open}} & Qwen2.5-7B & 0.618  & 0.830  & 0.710  & 0.256  & 0.399  & 0.439  & 0.739  & 0.105  & 0.649  & 0.324  & 0.659  & 0.517  \\
          & Qwen2.5-14B & 0.531  & 0.852  & 0.670  & 0.265  & 0.404  & 0.392  & 0.723  & 0.065  & 0.625  & 0.298  & 0.643  & 0.484  \\
          & Qwen2.5-32B & 0.602  & 0.865  & 0.716  & 0.263  & 0.421  & \underline{0.484}  & \underline{0.788}  & 0.090  & 0.646  & 0.346  & 0.687  & 0.531  \\
          & Qwen2.5-72B & \underline{0.655}  & \underline{0.867}  & \underline{0.747}  & 0.304  & 0.439  & 0.475  & 0.772  & \textbf{0.170} & \textbf{0.675} & \underline{0.369}  & \underline{0.691}  & \underline{0.558}  \\
          & Llama3.1-8B & 0.565  & 0.831  & 0.680  & \underline{0.321}  & \underline{0.441}  & 0.458  & 0.770  & \underline{0.120}  & 0.662  & 0.359  & 0.686  & 0.520  \\
          & Llama3.1-70B & \textbf{0.755} & \textbf{0.879} & \textbf{0.809} & \textbf{0.356} & \textbf{0.483} & \textbf{0.509} & \textbf{0.807} & 0.100  & \underline{0.673}  & \textbf{0.394} & \textbf{0.715} & \textbf{0.601} \\
    \midrule
    \multicolumn{1}{c}{\multirow{5}[2]{*}{Med}} & HuatuoGPT-o1-7B & 0.686  & 0.829  & 0.748  & 0.235  & 0.374  & 0.253  & 0.565  & 0.015  & 0.403  & 0.211  & 0.483  & 0.480  \\
          & HuatuoGPT-o1-8B & 0.509  & 0.839  & 0.652  & 0.281  & 0.374  & 0.159  & 0.522  & 0.015  & 0.441  & 0.181  & 0.473  & 0.416  \\
          & HuatuoGPT-o1-72B & \underline{0.689}  & 0.863  & \underline{0.764}  & \underline{0.338}  & 0.447  & \underline{0.263}  & 0.602  & 0.040  & 0.455  & 0.257  & 0.530  & 0.510  \\
          & HuatuoGPT-o1-70B & \textbf{0.699} & \underline{0.872}  & \textbf{0.774} & \textbf{0.365} & \textbf{0.470} & 0.256  & \underline{0.605}  & \underline{0.055}  & \underline{0.508}  & \underline{0.265}  & \underline{0.553}  & \underline{0.520}  \\
          & Baichuan-M1-14B & 0.651  & \textbf{0.879} & 0.750  & 0.327  & \underline{0.457}  & \textbf{0.463} & \textbf{0.783} & \textbf{0.140} & \textbf{0.644} & \textbf{0.366} & \textbf{0.688} & \textbf{0.558} \\
    \midrule
    \multicolumn{1}{c}{\multirow{6}[2]{*}{Reason}} & DsR1D-Qwen-7B & 0.250  & 0.661  & 0.428  & 0.201  & 0.324  & 0.232  & 0.573  & 0.019  & 0.438  & 0.189  & 0.491  & 0.309  \\
          & DsR1D-Llama-8b & 0.373  & 0.780  & 0.549  & 0.252  & 0.369  & 0.269  & 0.620  & 0.035  & 0.527  & 0.228  & 0.552  & 0.389  \\
          & DsR1D-Qwen-14B & 0.644  & 0.846  & 0.732  & 0.292  & 0.423  & 0.375  & 0.709  & 0.071  & 0.570  & 0.300  & 0.620  & 0.516  \\
          & DsR1D-Qwen-32B & 0.695  & 0.866  & 0.769  & 0.315  & 0.446  & \underline{0.385}  & \underline{0.718}  & \underline{0.103}  & \underline{0.640}  & \underline{0.318}  & \underline{0.652}  & \underline{0.544}  \\
          & DsR1D-Llama-70B & \underline{0.699}  & \textbf{0.889} & \underline{0.782}  & \underline{0.347}  & \underline{0.468}  & 0.338  & 0.692  & 0.063  & 0.593  & 0.300  & 0.626  & 0.541  \\
          & QwQ-32B & \textbf{0.725} & \underline{0.867}  & \textbf{0.787} & \textbf{0.412} & \textbf{0.529} & \textbf{0.511} & \textbf{0.807} & \textbf{0.109} & \textbf{0.656} & \textbf{0.416} & \textbf{0.716} & \textbf{0.601} \\
    \bottomrule
    \end{tabular}%
  }
  \label{tab:3}%
  \vspace{-6pt}
  \end{table}%

\textbf{Subjective Questions}.
The evaluation metric for open-ended diagnostic cases is defined as the clinical recall rate, which is calculated as the percentage of diseases correctly identified by the LLM.
To address the challenges of near-homograph and synonym matching while reducing the labor-intensive physician evaluation process, we use the LLMs for automated assessment.
For a clinical case containing $n$ diseases $[d_1, d_2, ..., d_n]$, we need to evaluate whether the prediction $\hat{y}$ of the model $M$ under assessment adequately covers each entity of the disease.
The workflow is as follows. First, the prediction $\hat{y}$ and each disease $d_i$ are input into a strict evaluation model $S$, which determines whether $\hat{y}$ includes $d_i$ by considering six dimensions: anatomical specificity, pathological mechanism, etiological tracing, temporal characteristics, laboratory markers, and therapeutic strategy.
Second, to mitigate potential false negatives arising from the rigid rule-based implementation of $S$, a secondary verification by a model $C$ is triggered when $S$ returns non-coverage. 
If $C$ finds that the disease is covered, the initial assessment of $S$ is replaced.
The prompts used for evaluation are provided in \textbf{Appendix \ref{appendix1.2}}.
Compared to using only one single evaluation model, our method reduces both false positives and false negatives in the results.

In implementaiton, we employ DeepSeek-V3-0324\cite{dsv3-0324} as the strict evaluation model and Claude-3.7-Sonnet \cite{claude3.7} as the check model. This method achieves agreement rates of 97.5\% and 97.7\% with the physicians in 393 Chinese and 132 English diseases, respectively.
Considering the comorbidity of multiple conditions within clinical cases, we perform analyses at two distinct levels: disease-specific recall and patient-level recall. The latter criterion is satisfied only when the model captures the full spectrum of concurrent diagnoses for a patient.
We employ four distinct prompts (see \textbf{Appendix \ref{appendix1.3}}) to mitigate the input sensitivity of LLMs: 
unconstrained context and output format, role-playing context with unstructured output, zero-shot chain-of-thought with unstructured output, and context-free with JSON-structured output. The results are averaged across all prompts.

\section{Experiments}

\subsection{Experiment Setup}
\vspace{-3pt}

We evaluate 26 LLMs on the ClinBench-HPB benchmark, encompassing four categories.

\textbf{Commercial LLMs:} GPT-4o-20240806 \cite{achiam2023gpt-4}, Claude3.5-sonnet-20241022 \cite{claude3.5}, DeepSeek-v3-20241226 \cite{deepseekv3}, DeepSeek-R1 \cite{deepseekr1}, Qwen2.5-Max \cite{qwen-max}, DeepSeek-v3-20250324 \cite{dsv3-0324}, OpenAI-o1-20241217 \cite{o1}, OpenAI-o3-mini-20250131 \cite{o3mini}, Gemini2.5-pro-20250325 \cite{gemini2.5}. We evaluate these models' efficacy in addressing real-world HPB diagnostic cases.


\textbf{Open-source General-purpose LLMs:} Qwen2.5-7B/14B/32B/72B-Instruct \cite{qwen2.5}, LLaMA3.1-8B/70B-Instruct \cite{llama3.1}. These models represent mainstream open-source LLMs that are commonly used as base models for fine-tuning. 

\textbf{Leading Medical LLMs:} HuatuoGPT-o1-7B/8B/70B/72B \cite{huatuo-o1} and Baichuan-m1-14B \cite{m1}. 
These top-performing medical LLMs are evaluated to determine their efficacy in HPB-related tasks compared to general-purpose models.

\textbf{Reasoning-Enhanced LLMs:} DeepSeek-R1-Dstill-Qwen7B/Llama8B/Qwen14B/Qwen32B/Llama70B \cite{deepseekr1}, QwQ-32B \cite{qwq32b}. 
These models have been specifically enhanced by mathematical and coding reasoning tasks, exhibiting superior reasoning capabilities compared to their base counterparts. 
We investigate whether such reasoning improvements can be directly transferred to medical applications.

We utilize official APIs for the commercial models, and deploy models in the other three categories locally on four H20 GPUs using vLLM \cite{vllm} for accelerated inference. 
For models incompatible with vLLM, we employ the officially recommended inference approach.
Additional implementation details are provided in \textbf{Appendix \ref{appendix2}}.

\subsection{Evaluation of Different LLMs}
\vspace{-3pt}

We evaluate the 26 LLMs on the proposed ClinBench-HPB. The results are presented in Table \ref{tab:3}.
The standard deviations across repeated experiments and the results for each prompt are presented in \textbf{Appendix \ref{appendix3}}.
All models are evaluated in the five subsets of the two tasks: multiple-choice question answering and clinical case diagnosis. 
The evaluation subsets comprised Chinese question-answering (CN-QA), English question-answering (EN-QA), journal-sourced cases (Journal), web-based medical cases (Website), and hospital-collected cases (Hospital).
Through within-group and between-group model comparison, we have several important findings.

\textbf{Commercial LLMs fail in real-world HPB diagnosis.}
As shown in Table \ref{tab:3}, commercial LLMs with hundreds of billions of parameters demonstrate outstanding performance on multiple-choice questions. The best-performing model can achieve an average accuracy of 0.84, which is sufficient to pass the medical license tests.
However, in real-world cases with more contextual information and complex clinical conditions, all models exhibit significant performance degradation, particularly on the Hospital subset.
Although some LLMs achieve acceptable performance at the disease level, their notably low patient-level recall indicates a substantial underdiagnosis problem. 
This is clinically unacceptable, as it can lead to significant treatment bias in practice.

\textbf{Open-source LLMs are approaching commercial models.}
Previous evaluations of LLMs in HPB medicine focus on commercial models, with limited assessment on open-source general-purpose models. 
As shown in Table \ref{tab:3}, open-source LLMs are reducing the performance gap with proprietary models. 
For example, Llama3.1-70B and DeepSeekV3-1226 show comparable performance (0.601 vs. 0.595), while surpassing proprietary models including GPT-4o, Claude 3.5 Sonnet, and Qwen2.5-Max in aggregate metrics.
Among open-source families, the similar-scale Llama3.1 performs slightly better than Qwen2.5 (Llama-8B vs. Qwen-7B, Llama-70B vs. Qwen-72B), while Qwen2.5-72B surpasses Llama-70B on the hospital subset.
These findings indicate that open-source models have acquired substantial HPB knowledge but require further improvement in analyzing complex tasks.


\begin{figure}
    \centering
    \includegraphics[width=1\linewidth]{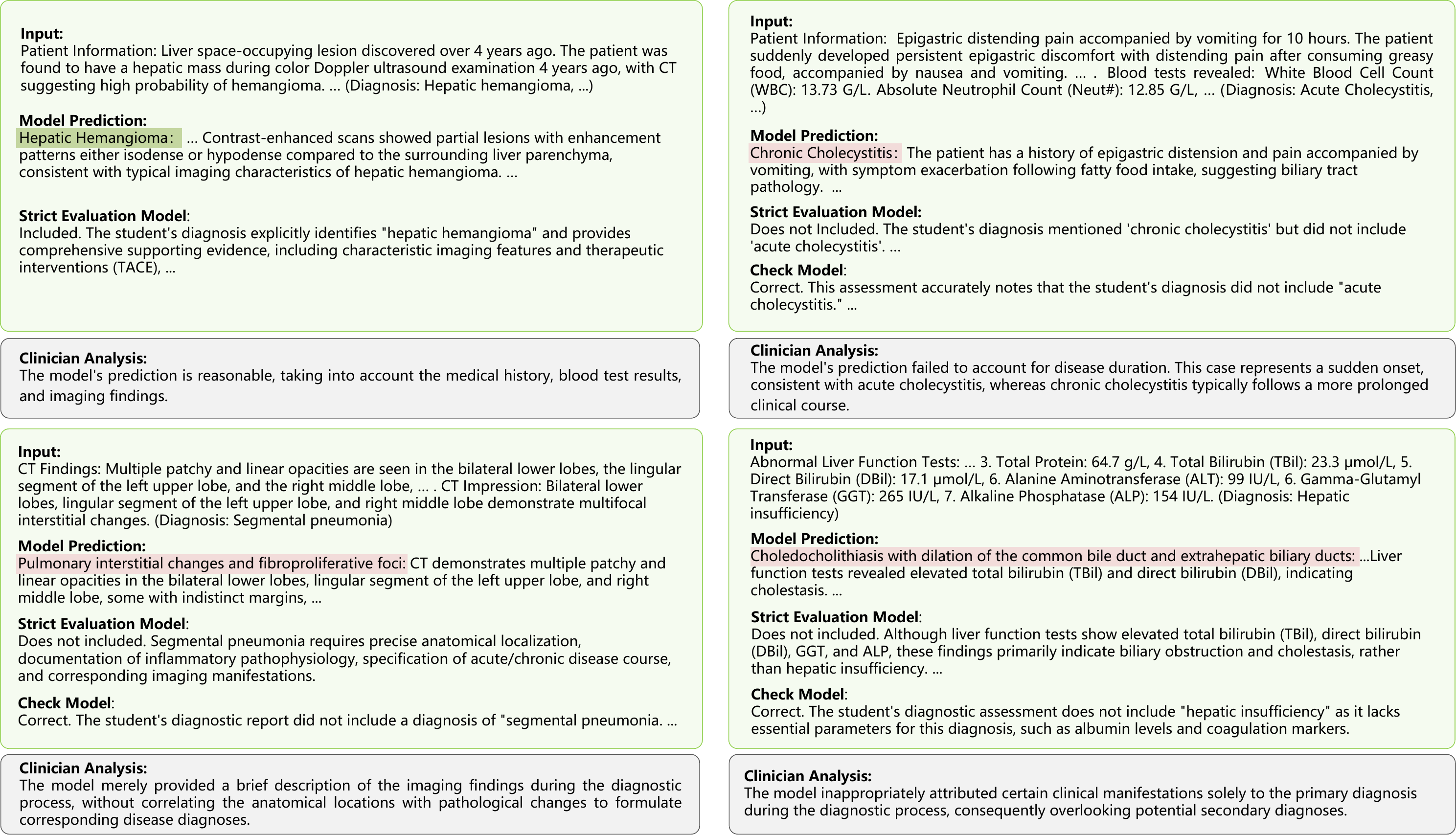}
    \caption{Examples of response generation for DeepSeekV3-1226 \cite{deepseekv3} on the Hospital subset.}
    \label{fig:3}
\end{figure}

\textbf{Medical LLMs exhibit performance degradation on OOD data.}
The training set of HuatuoGPT-o1 \cite{huatuo-o1} includes MedQA \cite{medqa} and MedMCQA \cite{medmcqa}, and its test sets are incorporated in our ClinBench-HPB.
Although this overlap of the test set could potentially benefit HuatuoGPT-o1, it does not demonstrate notable superiority over the baseline in overall performance because our ClinBench-HBP contains extensive multiple-choice questions from various sources. 
In clinical diagnostic tasks (\textit{e.g.}, on the Website and Hospital subsets), as presented in Table \ref{tab:3}, all the four HuatuoGPT-o1 models underperform their base models, suggesting the risk of overfitting to the training set.
One mitigation strategy can be large-scale medical knowledge injection, as evidenced by the fact that Baichuan-M1 \cite{m1} outperforms both the general-purpose model (Qwen-14B) and its reasoning-enhanced variant (DsR1D-Qwen14B) on all evaluation subsets.
However, this will incur substantial training costs and the model's performance on clinical diagnosis tasks remains suboptimal for practical applications.


\textbf{Reasoning-enhanced LLMs show promise in clinical applications.}
Guo et al. \cite{deepseekr1} applied supervised fine-tuning to LLMs using math and code data with reasoning chains, and demonstrated performance improvements in the corresponding downstream tasks.
On our benchmark, the smaller models (DsR1D-Qwen-7B/Llama-8B) exhibit limited capabilities, underperforming in multiple-choice tasks compared to both general-purpose models (Qwen2.5-7B \cite{qwen2.5}/Llama3.1-8B \cite{llama3.1}) and medical models (HuatuoGPT-o1-7B/8B \cite{huatuo-o1}).
However, as the model scale increases, DsR1D-Qwen-14B/32B demonstrate superior performance over baseline models across multiple subsets, as evidenced in Table \ref{tab:3}.
Notably, in diagnosis tasks, we observe that reasoning-enhanced models fine-tuned without medical data, relying solely on the base models' medical capability, surpass medical models on Website and Hospital subsets (DsR1D-Llama70B vs. HuatuoGPT-o1-70B).
This suggests that the reasoning capabilities elicited through math and code-based fine-tuning exhibit transfer potential to other scenarios, which could help mitigate data scarcity in medical applications.
Furthermore, due to its improved reasoning capabilities, QWQ-32B \cite{qwq32b} achieves comparable average performance to LLaMA3.1-70B, demonstrating that effective reasoning mechanisms can compensate for model size disadvantages in clinical case diagnosis tasks.

\subsection{Diagnostic Analysis}
\vspace{-3pt}

To investigate the issue of LLMs in practical diagnosis, we analyze the predictions of DeepSeekV3-1226 \cite{deepseekv3} in the Hospital subset, as shown in Fig. \ref{fig:3}.
More complete results are presented in \textbf{Appendix \ref{appendix4}}. 
The model demonstrates high diagnostic accuracy for several common HPB diseases, including hepatic hemangioma, cholelithiasis, liver cirrhosis, and hepatic cysts, which exhibit distinct features or imaging manifestations. In some cases, radiologists even provide the disease names in the "Impression" section of the imaging studies, resulting in relatively low analytical complexity.
In contrast, the model exhibits suboptimal performance in diagnosing diseases requiring the differentiation of temporal characteristics. 
As presented in Fig. \ref{fig:3}, when diagnosing acute cholecystitis, the model fails to recognize the short disease onset time in this case, resulting in misclassification as chronic cholecystitis.
Another frequent cause of diagnostic errors is the lack of in-depth analysis of the examination findings.
Segmental pneumonia is a common comorbidity among elderly patients, the diagnosis of which requires reasoning based on pulmonary segment descriptions, whereas in this case the model merely replicates the radiological observations.
Furthermore, certain laboratory findings can contain diagnostic information for multiple conditions, and insufficient analytical processing by the model could lead to missed diagnoses, as exemplified by the case of hepatic insufficiency.

\section{Conclusion and Limitation}

In this paper, we introduced ClinBench-HPB, a clinically oriented benchmark designed to assess the knowledge and practical diagnostic capabilities of LLMs in HPB diseases.
The benchmark comprised 3,535 multiple-choice questions and 337 real-world diagnostic cases. 
Through comprehensive benchmarking of 26 LLMs from commercial, open-source general-purpose, medical and reasoning-enhanced categories, we demonstrated that while commercial models performed well on multiple-choice questions, they still faced great challenges in open-ended real-world case diagnosis tasks.
Compared to baseline models, medical LLMs suffered from OOD performance degradation, while large-scale reasoning models demonstrated potentials in clinical applications.
The evaluation results revealed critical limitations of current LLMs in HPB disorders, underscoring that future medical LLM research should prioritize clinical diagnosis rather than simplistic medical examination questions.

There are some limitations of ClinBench-HPB. 
First, the benchmark in its current form is restricted to the text-modality. In the future, we will consider introducing clinical imaging data to enable multimodality evalaution. Second, the open-ended question evaluation method in ClinBench-HPB is a little complex and computationally expensive, and it inherents the stochasticity of LLM-based assessment. More efficient yet effective evaluation methods will be designed in the future.

{

\small
\bibliography{neurips_2025}
\bibliographystyle{unsrt}

}

\newpage







\newpage

\appendix

\section{Appendix}
In the appendix, we provide the following materials:

\begin{itemize}
\item All prompts used in our study (referring to Section 3.2 in the main paper).
\item Additional implementation details (referring to Section 4.1 in the main paper). 
\item Complete results of different LLMs on ClinBench-HPB (referring to Section 4.2 in the main paper).
\item Complete case diagnosis examples (referring to Section 4.4 in the main paper).

\end{itemize}

\subsection{Prompts}
\subsubsection{Prompts for Objective Multiple-Choice Questions}
\label{appendix1.1}
\begin{figure}[h]
  \centering
  \includegraphics[width=\textwidth]{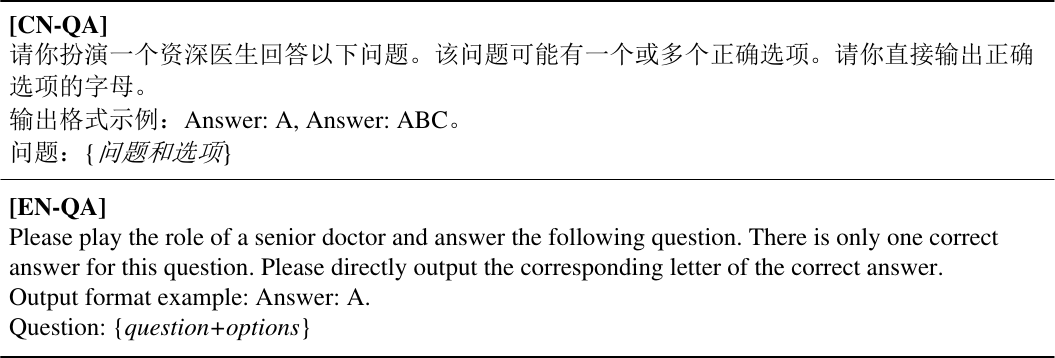}
  \caption{Illustration of the prompt templates for the objective multiple-choice questions.
 }
  \label{prompt:1}
\end{figure}

The prompts used in multiple-choice questions are shown in Fig. \ref{prompt:1}. The CN-QA subset includes both single-select and multiple-select questions, and the EN-QA subset consists exclusively of single-choice items.

\subsubsection{Prompts for LLM as a Judge}
\label{appendix1.2}
\begin{figure}
  \centering
  \includegraphics[width=\textwidth]{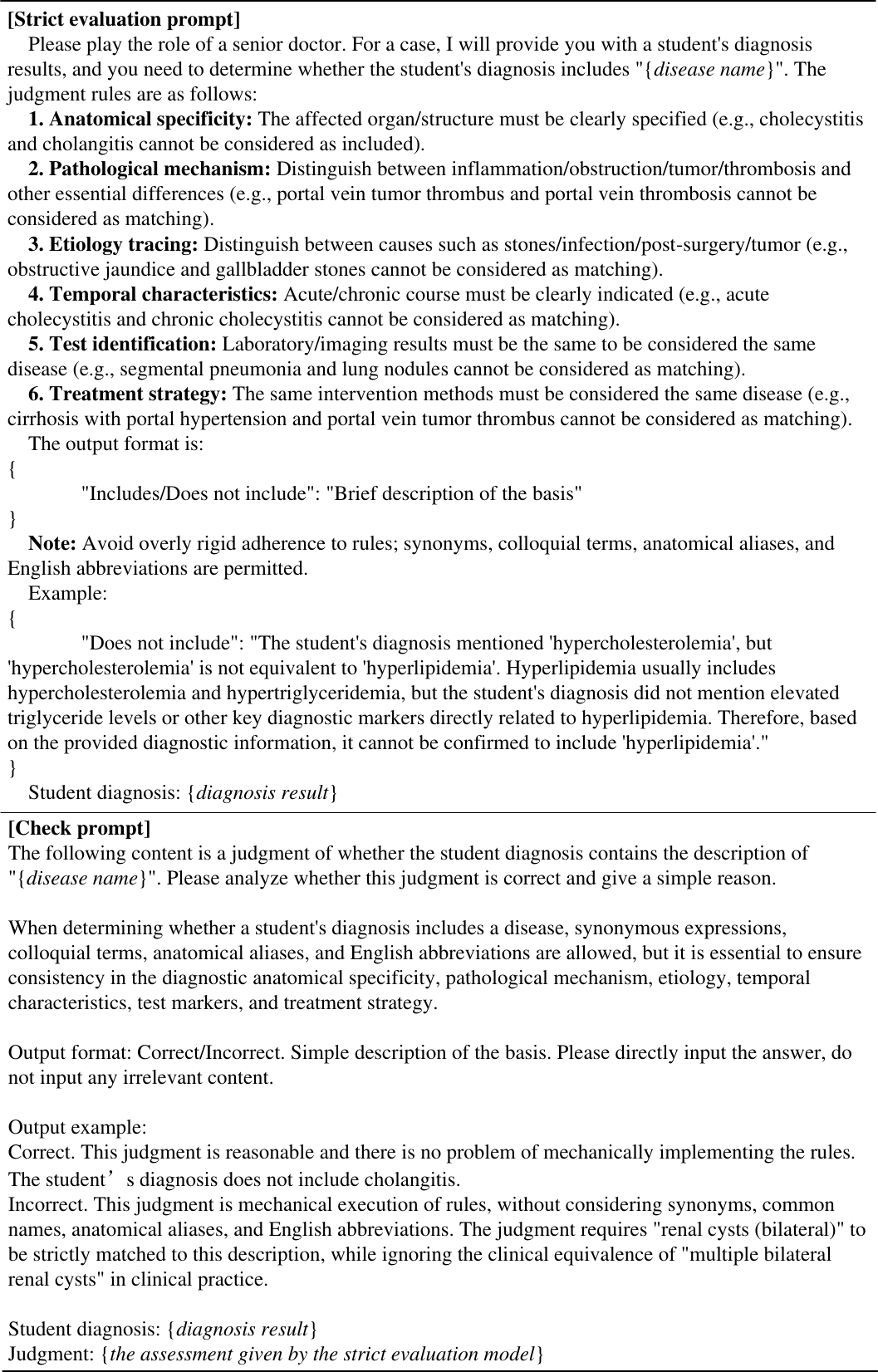}
  \caption{Illustration of the prompt templates for the strict evaluation model (top) and the check model (bottom).}
  \label{prompt:3}
\end{figure}

We first utilize DeepSeekV3-0324 \cite{dsv3-0324} as the strict evaluation model to assess whether the diagnosis result contains the reference disease. 
In addition, we employ Claude-3.7-sonnet \cite{claude3.7} as the check model to assess whether the strict evaluation model exhibits naive pattern matching that overlooks semantically equivalent expressions.
The prompt templates are presented in Fig. \ref{prompt:3}.

\subsubsection{Prompt Templates for Subjective Questions}
\label{appendix1.3}
\begin{figure}
  \centering
  \includegraphics[width=\textwidth]{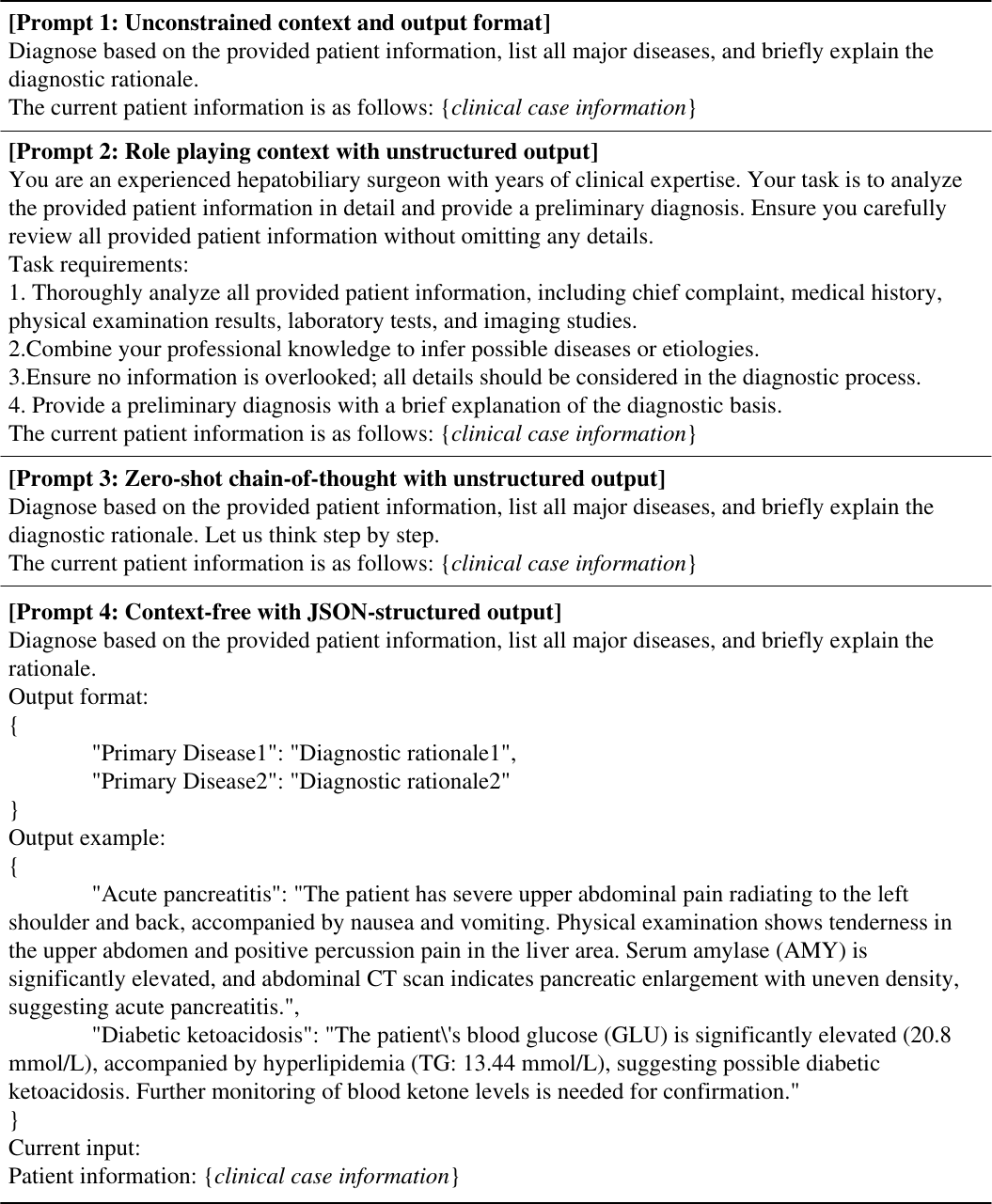}
  \caption{Illustration of the prompt templates for the subjective questions.} 
  \label{prompt:2}
\end{figure}

The prompts used in clinical diagnosis cases are shown in Fig. \ref{prompt:2}. We employ four distinct prompt templates to reduce the input sensitivity of the LLMs. 

\subsection{Additional Implementation Details}
\label{appendix2}
For locally deployed LLMs, we maintain each model's native top-p, top-k, and repetition penalty parameters. 
To ensure generation stability, we use temperature=0.6 for reasoning-enhanced models and DeepSeek-R1 with 4 iterations, followed by output averaging. For the other models, we apply single-pass greedy decoding.

For objective and subjective test items, option cyclic permutation and diverse prompting strategies are, respectively, employed to mitigate LLMs' sensitivity to input variations.
For instance, when processing a multiple-choice question with four options, reasoning-enhanced models and DeepSeek-R1 generate four responses per permutation (16 responses in total after all permutations). For a clinical diagnosis task, these models produce four responses for each of the four distinct prompts (16 responses in total).
Due to access limitations of OpenAI-o1 \cite{o1}, we randomly select 10\% of the questions from both CN-QA and EN-QA for evaluation, comprising 200 and 153 questions, respectively. The evaluation protocol for open-ended diagnosis is consistent with that of other models.

\subsection{Full Results of Model Comparison}
\label{appendix3}
For the multiple-choice questions in the CN-QA and EN-QA subsets, we perform cyclic right shifts on each option and repeat the predictions four times for both reasoning-enhanced models and DeepSeek-R1 under each condition. 
We compute the mean and standard deviation (SD) over multiple predictions for each question. The final results are obtained by averaging the means and SDs across all the questions in the subset, as shown in Table \ref{tab:4}.
The results of all LLMs across the Journal, Website, and Hospital subsets under the four distinct prompts are presented in Tables \ref{tab:5}, \ref{tab:6}, and \ref{tab:7}, respectively.

\begin{table}[htbp]
  \centering
  \caption{Results of different LLMs on the CN-QA and EN-QA subsets. "$\dagger$" means that the results are obtained on sampled subset.}
    \resizebox{0.8\textwidth}{!}{
    \begin{tabular}{cl|ccccc}
    \toprule
    \multicolumn{1}{c}{\multirow{3}[2]{*}{Category}} & \multicolumn{1}{c|}{\multirow{3}[2]{*}{Model}} & \multicolumn{5}{c}{Multiple Choice} \\
          &       & \multicolumn{2}{p{6.5em}}{CN-QA(2,000)} & \multicolumn{2}{p{6.5em}}{EN-QA(1,535)} & \multicolumn{1}{c}{\multirow{2}[1]{*}{$Avg$}} \\
          &       & \multicolumn{1}{c}{Mean} & \multicolumn{1}{c}{SD} & \multicolumn{1}{c}{Mean} & \multicolumn{1}{c}{SD} &  \\
    \midrule
    \multicolumn{1}{c}{\multirow{9}[2]{*}{Comm}} & GPT-4o \cite{achiam2023gpt-4} & 0.594  & 0.116  & 0.889  & 0.040  & 0.722  \\
          & OpenAI-o1 \cite{o1} & $0.760^\dagger$  & $0.056^\dagger$  & $0.917^\dagger$  & $0.024^\dagger$  & $0.828^\dagger$  \\
          & OpenAI-o3mini \cite{o3mini} & 0.666  & 0.092  & 0.894  & 0.026  & 0.765  \\
          & Claude3.5-sonnet\cite{claude3.5} & 0.623  & 0.127  & 0.878  & 0.056  & 0.734  \\
          & Gemini2.5-pro \cite{gemini2.5} & 0.724  & 0.075  & 0.897  & 0.032  & 0.799  \\
          & Qwen2.5-Max \cite{qwen-max} & 0.687  & 0.109  & 0.876  & 0.044  & 0.769  \\
          & DeepSeekV3-1226 \cite{deepseekv3} & 0.682  & 0.118  & 0.881  & 0.034  & 0.769  \\
          & DeepSeekV3-0324 \cite{dsv3-0324} & 0.803  & 0.077  & 0.887  & 0.041  & 0.839  \\
          & DeepSeek-R1 \cite{deepseekr1} & 0.799  & 0.095  & 0.893  & 0.043  & 0.840  \\
    \midrule
    \multicolumn{1}{c}{\multirow{6}[2]{*}{Open}} & Qwen2.5-7B \cite{qwen2.5} & 0.618  & 0.162  & 0.830  & 0.060  & 0.710  \\
          & Qwen2.5-14B \cite{qwen2.5} & 0.531  & 0.162  & 0.852  & 0.061  & 0.670  \\
          & Qwen2.5-32B \cite{qwen2.5} & 0.602  & 0.135  & 0.865  & 0.045  & 0.716  \\
          & Qwen2.5-72B \cite{qwen2.5} & 0.655  & 0.121  & 0.867  & 0.042  & 0.747  \\
          & Llama3.1-8B \cite{llama3.1} & 0.565  & 0.158  & 0.831  & 0.062  & 0.680  \\
          & Llama3.1-70B \cite{llama3.1} & 0.755  & 0.100  & 0.879  & 0.037  & 0.809  \\
    \midrule
    \multicolumn{1}{c}{\multirow{6}[2]{*}{Reason}} & DsR1D-Qwen-7B \cite{deepseekr1} & 0.250  & 0.292  & 0.661  & 0.246  & 0.428  \\
          & DsR1D-Llama-8b \cite{deepseekr1} & 0.373  & 0.311  & 0.780  & 0.177  & 0.549  \\
          & DsR1D-Qwen-14B \cite{deepseekr1} & 0.644  & 0.220  & 0.846  & 0.097  & 0.732  \\
          & DsR1D-Qwen-32B \cite{deepseekr1} & 0.695  & 0.198  & 0.866  & 0.088  & 0.769  \\
          & DsR1D-Llama-70B \cite{deepseekr1} & 0.699  & 0.192  & 0.889  & 0.062  & 0.782  \\
          & QwQ-32B \cite{qwq32b} & 0.725  & 0.238  & 0.867  & 0.075  & 0.787  \\
    \midrule
    \multicolumn{1}{c}{\multirow{5}[2]{*}{Med}} & HuatuoGPT-o1-7B \cite{huatuo-o1} & 0.686  & 0.134  & 0.829  & 0.079  & 0.748  \\
          & HuatuoGPT-o1-8B \cite{huatuo-o1} & 0.509  & 0.211  & 0.839  & 0.076  & 0.652  \\
          & HuatuoGPT-o1-72b \cite{huatuo-o1} & 0.689  & 0.126  & 0.863  & 0.062  & 0.764  \\
          & HuatuoGPT-o1-70b \cite{huatuo-o1} & 0.699  & 0.140  & 0.872  & 0.057  & 0.774  \\
          & Baichuan-M1-14B \cite{m1} & 0.651  & 0.165  & 0.879  & 0.047  & 0.750  \\
    \bottomrule
    \end{tabular}%
   }
  \label{tab:4}%
\end{table}%

\begin{table}[htbp]
  \centering
  \caption{Results of different LLMs on the Journal subset.}
    \resizebox{1.0\textwidth}{!}{
    \begin{tabular}{cl|cccccccccccc}
    \toprule
    \multicolumn{1}{c}{\multirow{3}[2]{*}{Category}} & \multicolumn{1}{c|}{\multirow{3}[2]{*}{Model}} & \multicolumn{12}{c}{Case Diagnosis-Journal} \\
          &       & \multicolumn{2}{c}{Prompt1} & \multicolumn{2}{c}{Prompt2} & \multicolumn{2}{c}{Prompt3} & \multicolumn{2}{c}{Prompt4} & \multirow{2}[1]{*}{$Avg_p$} & \multirow{2}[1]{*}{$SD_p$} & \multirow{2}[1]{*}{$Avg_d$} & \multirow{2}[1]{*}{$SD_d$} \\
          &       & Pt(120) & Dis(175) & Pt(120) & \multicolumn{1}{c}{Dis(175)} & Pt(120) & Dis(175) & Pt(120) & Dis(175) &       &       &       &  \\
    \midrule
    \multicolumn{1}{c}{\multirow{9}[2]{*}{Comm}} & GPT-4o \cite{achiam2023gpt-4} & 0.300  & 0.451  & 0.317  & \multicolumn{1}{c}{0.434} & 0.350  & 0.469  & 0.308  & 0.417  & 0.319  & 0.019  & 0.443  & 0.019  \\
          & OpenAI-o1 \cite{o1} & 0.517  & 0.611  & 0.492  & \multicolumn{1}{c}{0.571} & 0.508  & 0.617  & 0.442  & 0.549  & 0.490  & 0.029  & 0.587  & 0.028  \\
          & OpenAI-o3mini \cite{o3mini} & 0.592  & 0.674  & 0.617  & \multicolumn{1}{c}{0.669} & 0.575  & 0.669  & 0.450  & 0.537  & 0.558  & 0.064  & 0.637  & 0.058  \\
          & Claude3.5-sonnet\cite{claude3.5} & 0.467  & 0.566  & 0.250  & \multicolumn{1}{c}{0.406} & 0.450  & 0.560  & 0.408  & 0.526  & 0.394  & 0.086  & 0.514  & 0.065  \\
          & Gemini2.5-pro \cite{gemini2.5} & 0.500  & 0.611  & 0.517  & \multicolumn{1}{c}{0.634} & 0.525  & 0.611  & 0.458  & 0.554  & 0.500  & 0.026  & 0.603  & 0.030  \\
          & Qwen2.5-Max \cite{qwen-max} & 0.308  & 0.446  & 0.383  & \multicolumn{1}{c}{0.503} & 0.358  & 0.474  & 0.267  & 0.417  & 0.329  & 0.045  & 0.460  & 0.032  \\
          & DeepSeekV3-1226 \cite{deepseekv3} & 0.367  & 0.509  & 0.358  & \multicolumn{1}{c}{0.480} & 0.325  & 0.463  & 0.258  & 0.451  & 0.327  & 0.043  & 0.476  & 0.022  \\
          & DeepSeekV3-0324 \cite{dsv3-0324} & 0.367  & 0.503  & 0.392  & \multicolumn{1}{c}{0.514} & 0.442  & 0.571  & 0.325  & 0.463  & 0.381  & 0.042  & 0.513  & 0.039  \\
          & DeepSeek-R1 \cite{deepseekr1} & 0.440  & 0.564  & 0.440  & 0.559  & 0.419  & 0.549  & 0.433  & 0.540  & 0.433  & 0.009  & 0.553  & 0.009  \\
    \midrule
    \multicolumn{1}{c}{\multirow{6}[2]{*}{Open}} & Qwen2.5-7B \cite{qwen2.5} & 0.283  & 0.434  & 0.258  & \multicolumn{1}{c}{0.406} & 0.250  & 0.394  & 0.233  & 0.360  & 0.256  & 0.018  & 0.399  & 0.027  \\
          & Qwen2.5-14B \cite{qwen2.5} & 0.233  & 0.383  & 0.300  & \multicolumn{1}{c}{0.429} & 0.275  & 0.423  & 0.250  & 0.383  & 0.265  & 0.025  & 0.404  & 0.022  \\
          & Qwen2.5-32B \cite{qwen2.5} & 0.225  & 0.400  & 0.308  & \multicolumn{1}{c}{0.451} & 0.267  & 0.434  & 0.250  & 0.400  & 0.263  & 0.030  & 0.421  & 0.022  \\
          & Qwen2.5-72B \cite{qwen2.5} & 0.258  & 0.389  & 0.292  & \multicolumn{1}{c}{0.434} & 0.317  & 0.451  & 0.350  & 0.480  & 0.304  & 0.034  & 0.439  & 0.033  \\
          & Llama3.1-8B \cite{llama3.1} & 0.300  & 0.440  & 0.283  & \multicolumn{1}{c}{0.406} & 0.358  & 0.463  & 0.342  & 0.457  & 0.321  & 0.030  & 0.441  & 0.022  \\
          & Llama3.1-70B \cite{llama3.1} & 0.425  & 0.531  & 0.333  & 0.463  & 0.350  & 0.480  & 0.317  & 0.457  & 0.356  & 0.041  & 0.483  & 0.029  \\
    \midrule
    \multicolumn{1}{c}{\multirow{6}[2]{*}{Reason}} & DsR1D-Qwen-7B \cite{deepseekr1} & 0.196  & 0.316  & 0.206  & \multicolumn{1}{c}{0.354} & 0.188  & 0.323  & 0.213  & 0.304  & 0.201  & 0.010  & 0.324  & 0.019  \\
          & DsR1D-Llama-8b \cite{deepseekr1} & 0.269  & 0.374  & 0.242  & \multicolumn{1}{c}{0.376} & 0.242  & 0.367  & 0.254  & 0.359  & 0.252  & 0.011  & 0.369  & 0.007  \\
          & DsR1D-Qwen-14B \cite{deepseekr1} & 0.283  & 0.431  & 0.304  & \multicolumn{1}{c}{0.430} & 0.317  & 0.436  & 0.263  & 0.394  & 0.292  & 0.021  & 0.423  & 0.017  \\
          & DsR1D-Qwen-32B \cite{deepseekr1} & 0.298  & 0.444  & 0.323  & \multicolumn{1}{c}{0.436} & 0.317  & 0.459  & 0.323  & 0.447  & 0.315  & 0.010  & 0.446  & 0.008  \\
          & DsR1D-Llama-70B \cite{deepseekr1} & 0.375  & 0.486  & 0.346  & \multicolumn{1}{c}{0.470} & 0.356  & 0.480  & 0.310  & 0.436  & 0.347  & 0.024  & 0.468  & 0.019  \\
          & QwQ-32B \cite{qwq32b} & 0.413  & 0.527  & 0.410  & 0.530  & 0.408  & 0.533  & 0.417  & 0.524  & 0.412  & 0.003  & 0.529  & 0.003  \\
    \midrule
    \multicolumn{1}{c}{\multirow{5}[2]{*}{Med}} & HuatuoGPT-o1-7B \cite{huatuo-o1} & 0.275  & 0.406  & 0.242  & \multicolumn{1}{c}{0.366} & 0.250  & 0.394  & 0.175  & 0.331  & 0.235  & 0.037  & 0.374  & 0.029  \\
          & HuatuoGPT-o1-8B \cite{huatuo-o1} & 0.300  & 0.383  & 0.242  & \multicolumn{1}{c}{0.366} & 0.333  & 0.417  & 0.250  & 0.331  & 0.281  & 0.037  & 0.374  & 0.031  \\
          & HuatuoGPT-o1-72b \cite{huatuo-o1} & 0.300  & 0.434  & 0.375  & \multicolumn{1}{c}{0.469} & 0.292  & 0.406  & 0.383  & 0.480  & 0.338  & 0.042  & 0.447  & 0.029  \\
          & HuatuoGPT-o1-70b \cite{huatuo-o1} & 0.367  & 0.491  & 0.342  & \multicolumn{1}{c}{0.446} & 0.400  & 0.480  & 0.350  & 0.463  & 0.365  & 0.022  & 0.470  & 0.017  \\
          & Baichuan-M1-14B \cite{m1} & 0.358  & 0.497  & 0.317  & 0.440  & 0.300  & 0.457  & 0.333  & 0.434  & 0.327  & 0.022  & 0.457  & 0.025  \\
    \bottomrule
    \end{tabular}%
   }
  \label{tab:5}%
\end{table}%

\begin{table}[htbp]
  \centering
  \caption{Results of different LLMs on the Website subset.}
    \resizebox{1.0\textwidth}{!}{
    \begin{tabular}{cl|cccccccccccc}
    \toprule
    \multicolumn{1}{c}{\multirow{3}[2]{*}{Category}} & \multicolumn{1}{c|}{\multirow{3}[2]{*}{Model}} & \multicolumn{12}{c}{Case Diagnosis-Website} \\
          &       & \multicolumn{2}{c}{Prompt1} & \multicolumn{2}{c}{Prompt2} & \multicolumn{2}{c}{Prompt3} & \multicolumn{2}{c}{Prompt4} & \multirow{2}[1]{*}{$Avg_p$} & \multirow{2}[1]{*}{$SD_p$} & \multirow{2}[1]{*}{$Avg_d$} & \multirow{2}[1]{*}{$SD_d$} \\
          &       & Pt(167) & Dis(614) & Pt(167) & \multicolumn{1}{c}{Dis(614)} & Pt(167) & Dis(614) & Pt(167) & Dis(614) &       &       &       &  \\
    \midrule
    \multicolumn{1}{c}{\multirow{9}[2]{*}{Comm}} & GPT-4o \cite{achiam2023gpt-4} & 0.425  & 0.761  & 0.395  & \multicolumn{1}{c}{0.733} & 0.419  & 0.757  & 0.275  & 0.630  & 0.379  & 0.061  & 0.720  & 0.053  \\
          & OpenAI-o1 \cite{o1} & 0.533  & 0.826  & 0.581  & \multicolumn{1}{c}{0.819} & 0.677  & 0.889  & 0.311  & 0.673  & 0.525  & 0.134  & 0.802  & 0.079  \\
          & OpenAI-o3mini \cite{o3mini} & 0.461  & 0.796  & 0.413  & \multicolumn{1}{c}{0.739} & 0.778  & 0.909  & 0.503  & 0.774  & 0.539  & 0.142  & 0.805  & 0.064  \\
          & Claude3.5-sonnet\cite{claude3.5} & 0.641  & 0.876  & 0.461  & \multicolumn{1}{c}{0.792} & 0.659  & 0.871  & 0.425  & 0.772  & 0.546  & 0.104  & 0.828  & 0.047  \\
          & Gemini2.5-pro \cite{gemini2.5} & 0.587  & 0.847  & 0.485  & \multicolumn{1}{c}{0.803} & 0.539  & 0.839  & 0.407  & 0.749  & 0.504  & 0.067  & 0.809  & 0.039  \\
          & Qwen2.5-Max \cite{qwen-max} & 0.449  & 0.779  & 0.509  & \multicolumn{1}{c}{0.813} & 0.437  & 0.788  & 0.287  & 0.669  & 0.421  & 0.082  & 0.762  & 0.055  \\
          & DeepSeekV3-1226 \cite{deepseekv3} & 0.665  & 0.881  & 0.563  & \multicolumn{1}{c}{0.836} & 0.593  & 0.860  & 0.371  & 0.726  & 0.548  & 0.108  & 0.826  & 0.060  \\
          & DeepSeekV3-0324 \cite{dsv3-0324} & 0.599  & 0.855  & 0.473  & \multicolumn{1}{c}{0.775} & 0.623  & 0.857  & 0.413  & 0.751  & 0.527  & 0.087  & 0.809  & 0.047  \\
          & DeepSeek-R1 \cite{deepseekr1} & 0.624  & 0.860  & 0.482  & 0.767  & 0.626  & 0.859  & 0.316  & 0.654  & 0.512  & 0.127  & 0.785  & 0.084  \\
    \midrule
    \multicolumn{1}{c}{\multirow{6}[2]{*}{Open}} & Qwen2.5-7B \cite{qwen2.5} & 0.473  & 0.780  & 0.575  & \multicolumn{1}{c}{0.836} & 0.533  & 0.795  & 0.174  & 0.547  & 0.439  & 0.157  & 0.739  & 0.113  \\
          & Qwen2.5-14B \cite{qwen2.5} & 0.425  & 0.749  & 0.461  & \multicolumn{1}{c}{0.757} & 0.401  & 0.728  & 0.281  & 0.656  & 0.392  & 0.067  & 0.723  & 0.040  \\
          & Qwen2.5-32B \cite{qwen2.5} & 0.449  & 0.770  & 0.551  & \multicolumn{1}{c}{0.836} & 0.515  & 0.803  & 0.419  & 0.743  & 0.484  & 0.052  & 0.788  & 0.035  \\
          & Qwen2.5-72B \cite{qwen2.5} & 0.497  & 0.783  & 0.479  & \multicolumn{1}{c}{0.798} & 0.599  & 0.840  & 0.323  & 0.668  & 0.475  & 0.099  & 0.772  & 0.064  \\
          & Llama3.1-8B \cite{llama3.1} & 0.473  & 0.769  & 0.539  & \multicolumn{1}{c}{0.822} & 0.473  & 0.788  & 0.347  & 0.702  & 0.458  & 0.069  & 0.770  & 0.044  \\
          & Llama3.1-70B \cite{llama3.1} & 0.515  & 0.811  & 0.545  & 0.845  & 0.551  & 0.816  & 0.425  & 0.757  & 0.509  & 0.050  & 0.807  & 0.032  \\
    \midrule
    \multicolumn{1}{c}{\multirow{6}[2]{*}{Reason}} & DsR1D-Qwen-7B \cite{deepseekr1} & 0.260  & 0.625  & 0.272  & \multicolumn{1}{c}{0.601} & 0.289  & 0.643  & 0.108  & 0.423  & 0.232  & 0.073  & 0.573  & 0.088  \\
          & DsR1D-Llama-8b \cite{deepseekr1} & 0.289  & 0.655  & 0.301  & \multicolumn{1}{c}{0.656} & 0.326  & 0.676  & 0.160  & 0.495  & 0.269  & 0.064  & 0.620  & 0.073  \\
          & DsR1D-Qwen-14B \cite{deepseekr1} & 0.479  & 0.783  & 0.368  & \multicolumn{1}{c}{0.711} & 0.446  & 0.765  & 0.208  & 0.576  & 0.375  & 0.105  & 0.709  & 0.081  \\
          & DsR1D-Qwen-32B \cite{deepseekr1} & 0.478  & 0.797  & 0.412  & \multicolumn{1}{c}{0.740} & 0.449  & 0.778  & 0.204  & 0.558  & 0.385  & 0.108  & 0.718  & 0.095  \\
          & DsR1D-Llama-70B \cite{deepseekr1} & 0.428  & 0.761  & 0.307  & \multicolumn{1}{c}{0.664} & 0.383  & 0.735  & 0.234  & 0.606  & 0.338  & 0.074  & 0.692  & 0.061  \\
          & QwQ-32B \cite{qwq32b} & 0.572  & 0.844  & 0.524  & 0.812  & 0.533  & 0.821  & 0.416  & 0.752  & 0.511  & 0.058  & 0.807  & 0.034  \\
    \midrule
    \multicolumn{1}{c}{\multirow{5}[2]{*}{Med}} & HuatuoGPT-o1-7B \cite{huatuo-o1} & 0.317  & 0.609  & 0.240  & \multicolumn{1}{c}{0.560} & 0.305  & 0.619  & 0.150  & 0.471  & 0.253  & 0.067  & 0.565  & 0.059  \\
          & HuatuoGPT-o1-8B \cite{huatuo-o1} & 0.162  & 0.549  & 0.144  & \multicolumn{1}{c}{0.492} & 0.192  & 0.557  & 0.138  & 0.490  & 0.159  & 0.021  & 0.522  & 0.031  \\
          & HuatuoGPT-o1-72b \cite{huatuo-o1} & 0.299  & 0.642  & 0.228  & \multicolumn{1}{c}{0.552} & 0.281  & 0.640  & 0.246  & 0.575  & 0.263  & 0.028  & 0.602  & 0.040  \\
          & HuatuoGPT-o1-70b \cite{huatuo-o1} & 0.293  & 0.650  & 0.216  & \multicolumn{1}{c}{0.531} & 0.293  & 0.651  & 0.222  & 0.588  & 0.256  & 0.037  & 0.605  & 0.050  \\
          & Baichuan-M1-14B \cite{m1} & 0.497  & 0.811  & 0.533  & 0.818  & 0.527  & 0.826  & 0.293  & 0.676  & 0.463  & 0.099  & 0.783  & 0.062  \\
    \bottomrule
    \end{tabular}%
   }
  \label{tab:6}%
\end{table}%

\begin{table}[htbp]
  \centering
  \caption{Results of different LLMs on the Hospital subset.}
    \resizebox{1.0\textwidth}{!}{
    \begin{tabular}{cl|cccccccccccc}
    \toprule
    \multicolumn{1}{c}{\multirow{3}[2]{*}{Category}} & \multicolumn{1}{c|}{\multirow{3}[2]{*}{Model}} & \multicolumn{12}{c}{Case Diagnosis-Hospital} \\
          &       & \multicolumn{2}{c}{Prompt1} & \multicolumn{2}{c}{Prompt2} & \multicolumn{2}{c}{Prompt3} & \multicolumn{2}{c}{Prompt4} & \multirow{2}[1]{*}{$Avg_p$} & \multirow{2}[1]{*}{$SD_p$} & \multirow{2}[1]{*}{$Avg_d$} & \multirow{2}[1]{*}{$SD_d$} \\
          &       & Pt(50) & Dis(393) & Pt(50) & \multicolumn{1}{c}{Dis(393)} & Pt(50) & Dis(393) & Pt(50) & Dis(393) &       &       &       &  \\
    \midrule
    \multicolumn{1}{c}{\multirow{9}[2]{*}{Comm}} & GPT-4o \cite{achiam2023gpt-4} & 0.120  & 0.634  & 0.080  & \multicolumn{1}{c}{0.588} & 0.100  & 0.621  & 0.060  & 0.550  & 0.090  & 0.022  & 0.598  & 0.033  \\
          & OpenAI-o1 \cite{o1} & 0.180  & 0.710  & 0.180  & \multicolumn{1}{c}{0.710} & 0.180  & 0.710  & 0.180  & 0.710  & 0.180  & 0.000  & 0.710  & 0.000  \\
          & OpenAI-o3mini \cite{o3mini} & 0.140  & 0.664  & 0.120  & \multicolumn{1}{c}{0.585} & 0.100  & 0.700  & 0.100  & 0.613  & 0.115  & 0.017  & 0.641  & 0.044  \\
          & Claude3.5-sonnet\cite{claude3.5} & 0.100  & 0.557  & 0.000  & \multicolumn{1}{c}{0.509} & 0.020  & 0.580  & 0.080  & 0.634  & 0.050  & 0.041  & 0.570  & 0.045  \\
          & Gemini2.5-pro \cite{gemini2.5} & 0.140  & 0.728  & 0.180  & \multicolumn{1}{c}{0.779} & 0.160  & 0.730  & 0.060  & 0.613  & 0.135  & 0.046  & 0.712  & 0.061  \\
          & Qwen2.5-Max \cite{qwen-max} & 0.080  & 0.667  & 0.260  & \multicolumn{1}{c}{0.763} & 0.120  & 0.728  & 0.060  & 0.590  & 0.130  & 0.078  & 0.687  & 0.066  \\
          & DeepSeekV3-1226 \cite{deepseekv3} & 0.320  & 0.819  & 0.240  & \multicolumn{1}{c}{0.746} & 0.280  & 0.799  & 0.040  & 0.595  & 0.220  & 0.108  & 0.740  & 0.088  \\
          & DeepSeekV3-0324 \cite{dsv3-0324} & 0.240  & 0.771  & 0.100  & \multicolumn{1}{c}{0.664} & 0.260  & 0.779  & 0.100  & 0.651  & 0.175  & 0.075  & 0.716  & 0.059  \\
          & DeepSeek-R1 \cite{deepseekr1} & 0.175  & 0.719  & 0.085  & 0.630  & 0.145  & 0.719  & 0.025  & 0.522  & 0.108  & 0.058  & 0.648  & 0.081  \\
    \midrule
    \multicolumn{1}{c}{\multirow{6}[2]{*}{Open}} & Qwen2.5-7B \cite{qwen2.5} & 0.140  & 0.687  & 0.140  & \multicolumn{1}{c}{0.710} & 0.120  & 0.692  & 0.020  & 0.506  & 0.105  & 0.050  & 0.649  & 0.083  \\
          & Qwen2.5-14B \cite{qwen2.5} & 0.040  & 0.636  & 0.060  & \multicolumn{1}{c}{0.618} & 0.100  & 0.656  & 0.060  & 0.588  & 0.065  & 0.022  & 0.625  & 0.025  \\
          & Qwen2.5-32B \cite{qwen2.5} & 0.060  & 0.641  & 0.140  & \multicolumn{1}{c}{0.695} & 0.120  & 0.664  & 0.040  & 0.585  & 0.090  & 0.041  & 0.646  & 0.040  \\
          & Qwen2.5-72B \cite{qwen2.5} & 0.200  & 0.715  & 0.180  & \multicolumn{1}{c}{0.720} & 0.200  & 0.730  & 0.100  & 0.534  & 0.170  & 0.041  & 0.675  & 0.081  \\
          & Llama3.1-8B \cite{llama3.1} & 0.100  & 0.626  & 0.160  & \multicolumn{1}{c}{0.702} & 0.100  & 0.651  & 0.120  & 0.669  & 0.120  & 0.024  & 0.662  & 0.028  \\
          & Llama3.1-70B \cite{llama3.1} & 0.100  & 0.654  & 0.120  & 0.702  & 0.100  & 0.687  & 0.080  & 0.649  & 0.100  & 0.014  & 0.673  & 0.022  \\
    \midrule
    \multicolumn{1}{c}{\multirow{6}[2]{*}{Reason}} & DsR1D-Qwen-7B \cite{deepseekr1} & 0.020  & 0.480  & 0.020  & \multicolumn{1}{c}{0.440} & 0.030  & 0.485  & 0.005  & 0.345  & 0.019  & 0.009  & 0.438  & 0.056  \\
          & DsR1D-Llama-8b \cite{deepseekr1} & 0.035  & 0.564  & 0.030  & \multicolumn{1}{c}{0.529} & 0.060  & 0.572  & 0.015  & 0.442  & 0.035  & 0.016  & 0.527  & 0.051  \\
          & DsR1D-Qwen-14B \cite{deepseekr1} & 0.130  & 0.656  & 0.065  & \multicolumn{1}{c}{0.580} & 0.080  & 0.620  & 0.010  & 0.424  & 0.071  & 0.043  & 0.570  & 0.089  \\
          & DsR1D-Qwen-32B \cite{deepseekr1} & 0.150  & 0.715  & 0.085  & \multicolumn{1}{c}{0.650} & 0.160  & 0.709  & 0.015  & 0.485  & 0.103  & 0.058  & 0.640  & 0.093  \\
          & DsR1D-Llama-70B \cite{deepseekr1} & 0.095  & 0.665  & 0.060  & \multicolumn{1}{c}{0.558} & 0.050  & 0.627  & 0.045  & 0.523  & 0.063  & 0.020  & 0.593  & 0.056  \\
          & QwQ-32B \cite{qwq32b} & 0.145  & 0.720  & 0.090  & 0.659  & 0.160  & 0.711  & 0.040  & 0.535  & 0.109  & 0.047  & 0.656  & 0.074  \\
    \midrule
    \multicolumn{1}{c}{\multirow{5}[2]{*}{Med}} & HuatuoGPT-o1-7B \cite{huatuo-o1} & 0.020  & 0.443  & 0.020  & \multicolumn{1}{c}{0.377} & 0.020  & 0.478  & 0.000  & 0.316  & 0.015  & 0.009  & 0.403  & 0.062  \\
          & HuatuoGPT-o1-8B \cite{huatuo-o1} & 0.020  & 0.486  & 0.000  & \multicolumn{1}{c}{0.382} & 0.020  & 0.448  & 0.020  & 0.448  & 0.015  & 0.009  & 0.441  & 0.038  \\
          & HuatuoGPT-o1-72b \cite{huatuo-o1} & 0.100  & 0.524  & 0.000  & \multicolumn{1}{c}{0.410} & 0.060  & 0.504  & 0.000  & 0.384  & 0.040  & 0.042  & 0.455  & 0.060  \\
          & HuatuoGPT-o1-70b \cite{huatuo-o1} & 0.060  & 0.550  & 0.040  & \multicolumn{1}{c}{0.417} & 0.060  & 0.557  & 0.060  & 0.506  & 0.055  & 0.009  & 0.508  & 0.056  \\
          & Baichuan-M1-14B \cite{m1} & 0.140  & 0.669  & 0.160  & 0.679  & 0.200  & 0.697  & 0.060  & 0.532  & 0.140  & 0.051  & 0.644  & 0.066  \\
    \bottomrule
    \end{tabular}%
   }
  \label{tab:7}%
\end{table}%

\newpage

\subsection{Examples of Complete Diagnosis Cases}
\label{appendix4}
\begin{figure}[h]
  \centering
  \includegraphics[width=\textwidth]{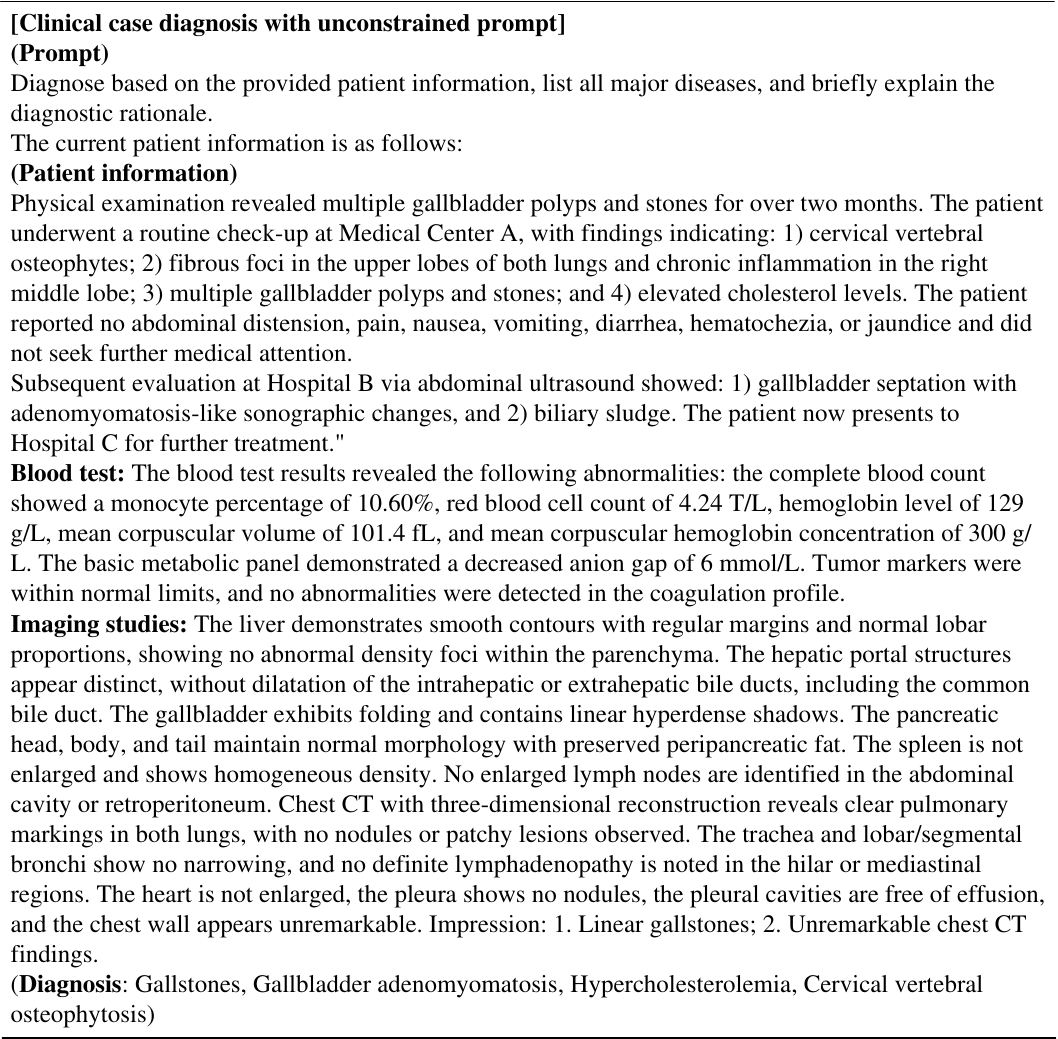}
  \caption{Clinical case 1 from the Hospital subset. The original Chinese text has been translated into the corresponding English version.}
  \label{prompt:4}
\end{figure}

\begin{figure}
  \centering
  \includegraphics[width=\textwidth]{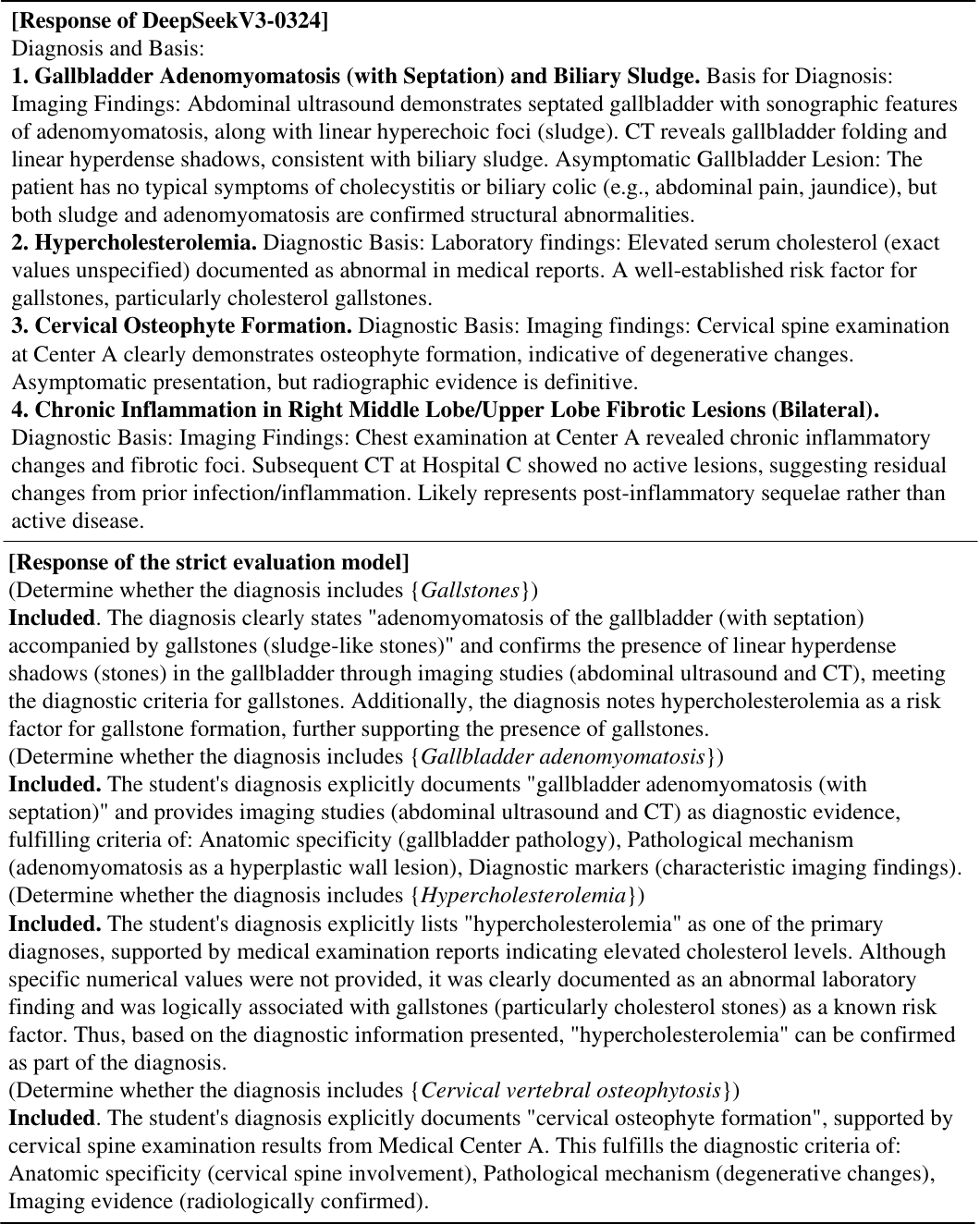}
  \caption{Diagnostic results of DeepSeekV3-0324 \cite{dsv3-0324} and its evaluation.}
  \label{prompt:5}
\end{figure}

\begin{figure}
  \centering
  \includegraphics[width=\textwidth]{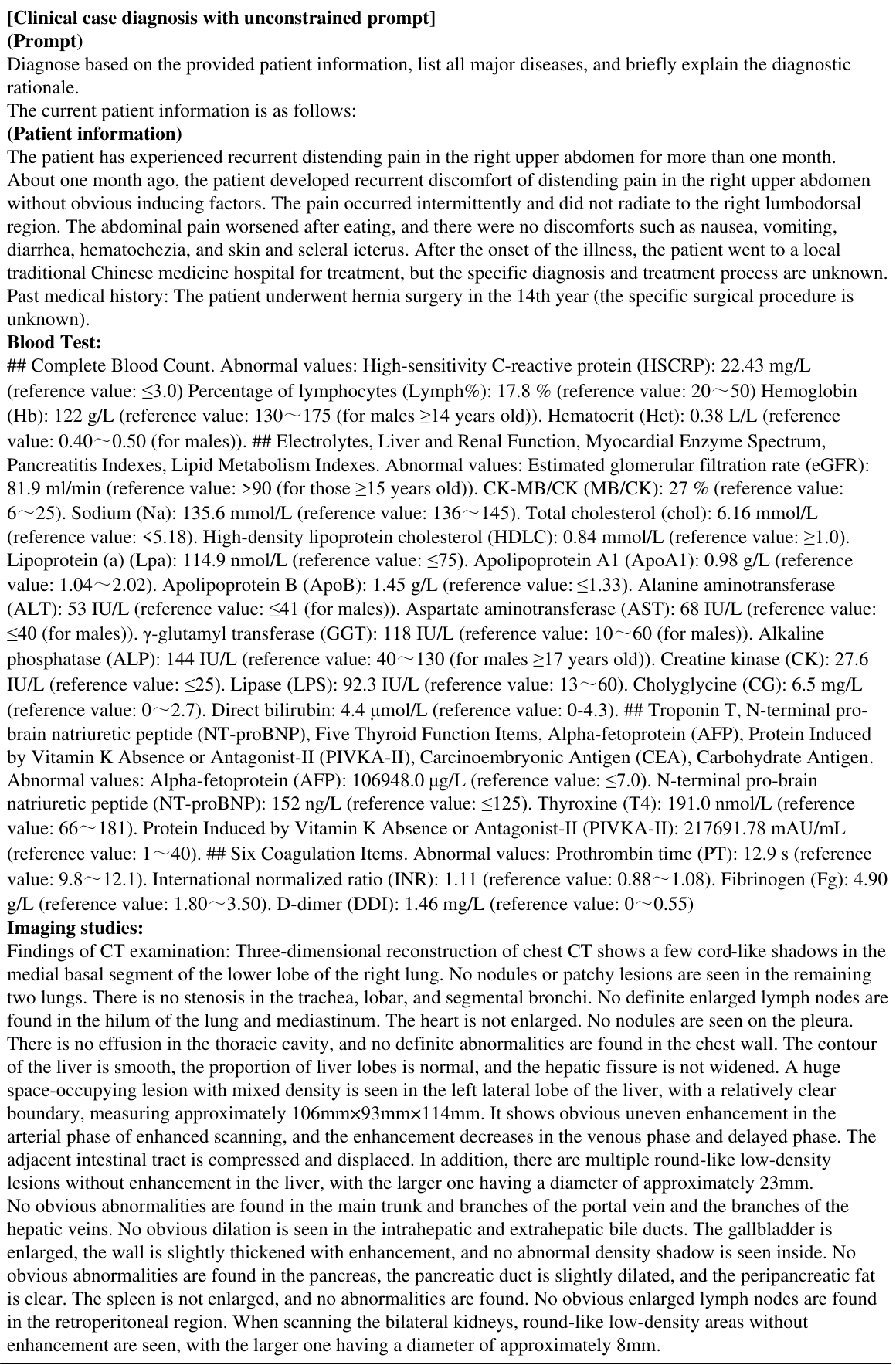}
  \caption{Clinical case 2 from the Hospital subset (Part 1).}
  \label{prompt:6}
\end{figure}

\begin{figure}
  \centering
  \includegraphics[width=\textwidth]{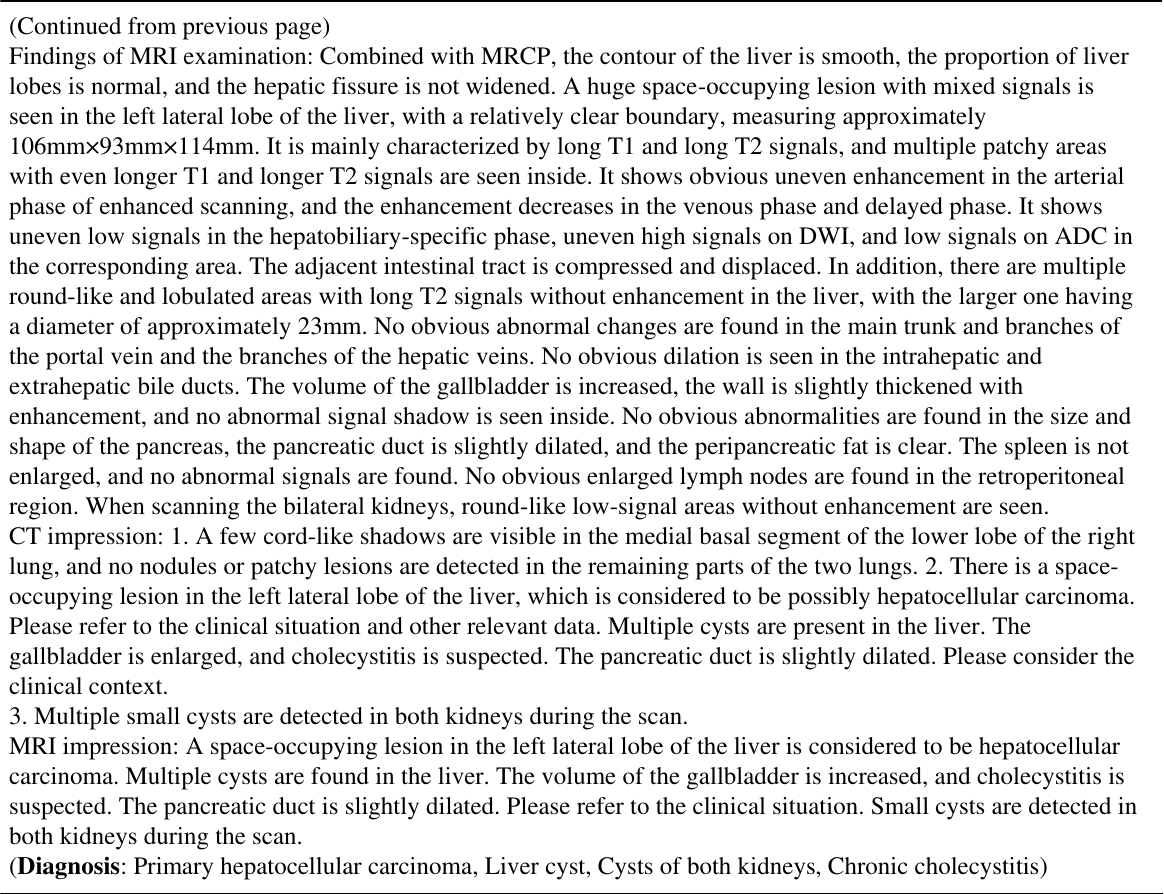}
  \caption{Clinical case 2 from the Hospital subset (Part 2).}
  \label{prompt:7}
\end{figure}

\begin{figure}
  \centering
  \includegraphics[width=\textwidth]{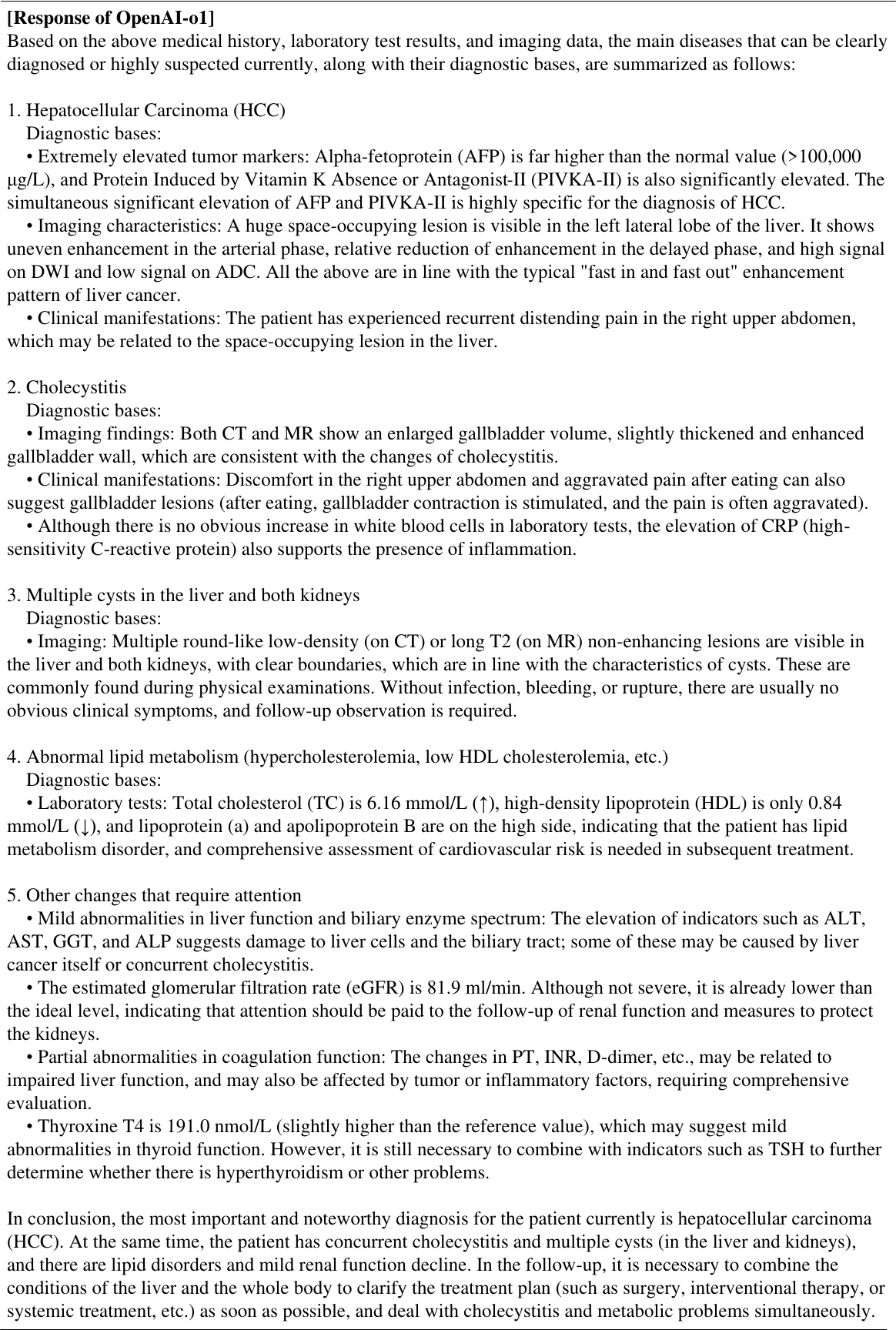}
  \caption{Diagnostic results of OpenAI-o1 \cite{o1}.}
  \label{prompt:8}
\end{figure}

\begin{figure}
  \centering
  \includegraphics[width=\textwidth]{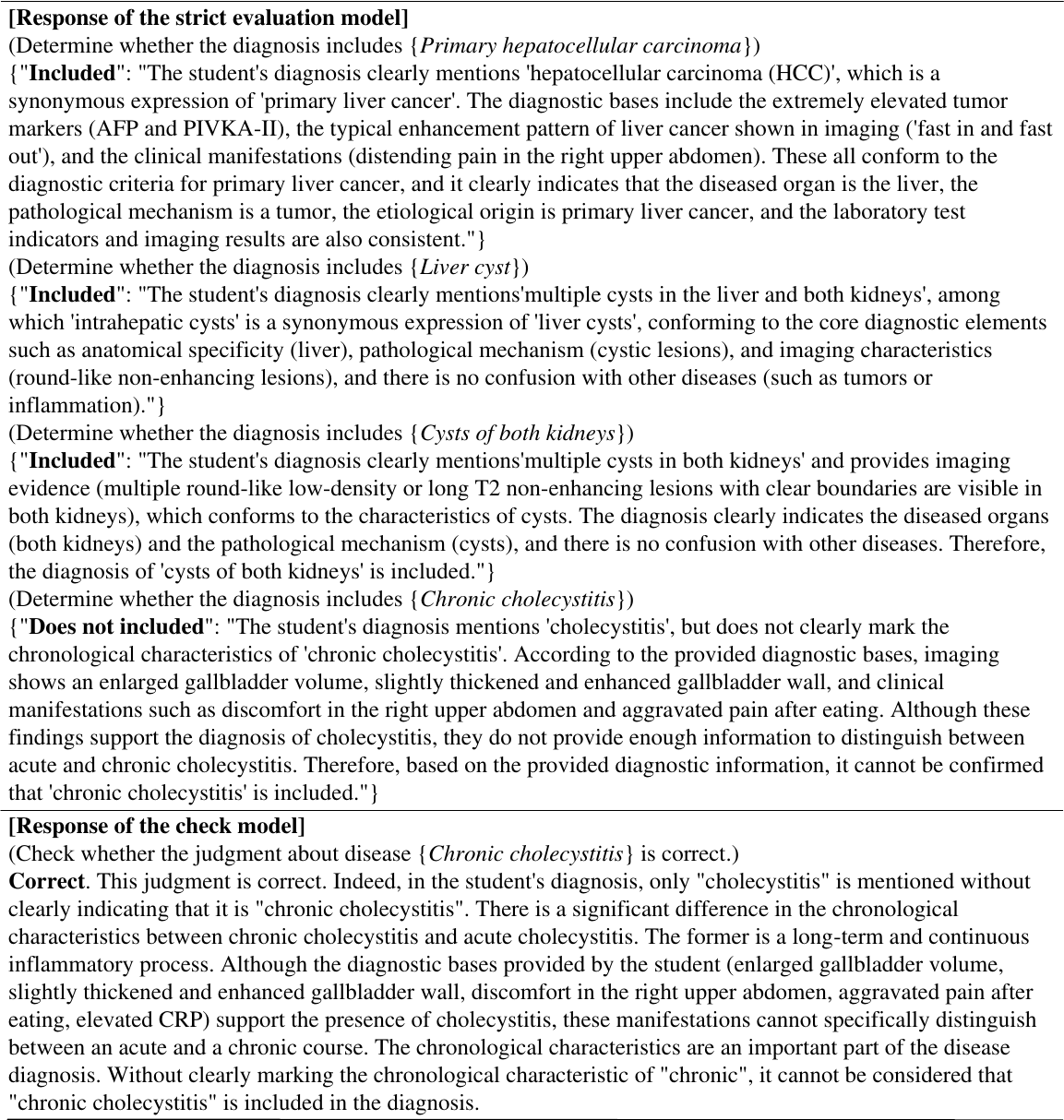}
  \caption{Evaluation of the prediction from OpenAI-o1 \cite{o1}.}
  \label{prompt:9}
\end{figure}

We present two complete diagnosis cases from the Hospital subset, demonstrating the diagnostic capability of DeepSeekV3-0324 \cite{dsv3-0324} and OpenAI-o1 \cite{o1}, respectively. 
The results are evaluated using the strict evaluation model and the check model. The two cases are illustrated in Figs. \ref{prompt:4}, \ref{prompt:5} and Figs. \ref{prompt:6}, \ref{prompt:7}, \ref{prompt:8}, \ref{prompt:9}, respectively.


\end{document}